\documentclass[amsmath,amssymb,aps,pra,reprint,superscriptaddress,nourl]{revtex4-1}
\usepackage{graphicx,hyperref,xcolor}
\usepackage{epstopdf}

\newcommand {\ket}[1] {\left|{#1}\right\rangle}
\newcommand {\bra}[1] {\langle{#1}|}

\begin{document}

%\preprint{}

\title{Quantum limit for two-dimensional resolution of two incoherent optical point sources}


\author{Shan Zheng Ang}
%\email[]{eleasz@nus.edu.sg}
%\thanks{}
\affiliation{Department of Electrical and Computer Engineering, National University of Singapore, 4 Engineering Drive 3, Singapore 117583}
\author{Ranjith Nair}
\affiliation{Department of Electrical and Computer Engineering, National University of Singapore, 4 Engineering Drive 3, Singapore 117583}
\author{Mankei Tsang}
\affiliation{Department of Electrical and Computer Engineering, National University of Singapore, 4 Engineering Drive 3, Singapore 117583}
\affiliation{Department of Physics, National University of Singapore, 2 Science Drive 3, Singapore 117551}


\date{\today}

\begin{abstract}
We obtain the multiple-parameter quantum Cram\'er-Rao bound for estimating the transverse Cartesian components of the centroid and separation of two incoherent optical point sources using an imaging system with finite spatial bandwidth. Under quite general and realistic assumptions on the point-spread function of the imaging system, and for weak source strengths, we show that the Cram\'er-Rao bounds for the $x$ and $y$ components of the separation are independent of the values of those components, which may be well below the conventional Rayleigh resolution limit. We also propose two linear optics-based measurement methods that approach the quantum bound for the estimation of the Cartesian components of the separation once the centroid has been located. One of the methods is an interferometric scheme that approaches the quantum bound for sub-Rayleigh separations. The other method using fiber coupling can in principle attain the bound regardless of the distance between the two sources.
\end{abstract}

%\keywords{}

\maketitle

\section{Introduction}
Rayleigh's criterion for resolution of two incoherent point sources \cite{rayleigh1879xxxi} has been the most widely accepted criterion for optical resolution since its formulation in 1879.
Being rooted in the optical measurement technology of its era, Rayleigh's criterion neglects the discrete and stochastic nature of the photodetection process. By adopting a stochastic framework, the studies~\cite{bettens99,vanaert02,ram06}  gave a modern formulation of the criterion for two sources radiating independently and incoherently. Using the Cram\'er-Rao (CR) bound of classical estimation theory~\cite{vantrees01}, they showed that the localization accuracy of any unbiased estimator based on image-plane photon counting deteriorates rapidly on approaching sub-Rayleigh separations. 

 In the past few decades, advances in far-field super-resolution techniques in microscopy \cite{betzig95,moerner89,hell94} (see ref.~\cite{WS15} for a review) have
enabled us to sidestep Rayleigh's limit. Still, as they require  that nearby sources are not emiting at the same time,  those technologies do not challenge Rayleigh's criterion fundamentally for independently emitting sources.

Very recently, the localization problem was reconsidered from the perspective of  quantum estimation theory using the quantum Cram\'er-Rao (QCR) bound~\cite{helstrom76,holevo11,paris09}. Following a preliminary study of the  fundamental localization limit for coherent sources in ref.~\cite{tsang15b}, Tsang \emph{et al.}~\cite{tsang16a} obtained the quantum limit on localizing two weak incoherent optical point sources in one-dimensional imaging.  Their quantum bound for estimating the separation between the sources  is independent of that separation and shows no deterioration when the two sources are closer than the conventional Rayleigh limit of the imaging system. Similar conclusions were reproduced using a semiclassical photodetection theory under a Poisson model~\cite{tsang16b}. In ref.~\cite{tsang16a}, a linear optical measurement -- SPADE (SPAtial-mode DEmultiplexing)-- was also proposed and shown to attain the QCR bound for any separation. Another measurement scheme --SLIVER (Super Localization by Image inVERsion interferometry) -- was  proposed in ref.~\cite{nair16} that approaches the QCR bound for sub-Rayleigh separations. 
Recent experimental work~\cite{SDL16,YTM+16,TFS17,PSH+16} inspired by the above proposals has substantiated the surprising findings of the above papers.

In further theoretical work, the QCR bound on the one-dimensional separation was calculated for thermal sources of arbitrary strength using a  Gaussian-state model, and variants of SPADE and SLIVER were shown to approach the quantum limit over arbitrarily large ranges of the separation  \cite{NT16Gaussian}. In ref.~\cite{lupo16}, the question of optimizing the quantum state of the source pair under an energy constraint was addressed and the optimum source states for sub-Rayleigh imaging were found. In ref.~\cite{RPS+16}, a systematic approach to finding optimal measurement modes in the image plane  for a given point-spread function was developed -- see also \cite{KGA17}. In \cite{Tsa17}, estimation of more general image parameters was studied, and in refs.~\cite{lu16,KGS16}, the resolution problem is addressed in terms of quantum detection theory.

In this paper, we address the problem of \emph{two-dimensional}, i.e., complete transverse-plane  localization of two incoherent optical point sources. Adopting the weak-source model of ref.~\cite{tsang16a}, we first obtain the full 4-parameter quantum Fisher information (QFI) matrix characterizing the ultimate precision of estimating all four transverse Cartesian coordinates  of the two sources. As in the one-dimensional case, the quantum bound suggests that the Rayleigh resolution limit is not fundamental and can be circumvented by an appropriate quantum measurement. We then focus on estimating the $x$- and $y$-components of the separation in the transverse plane once the centroid (midpoint) of the sources has been located.  Recent theoretical studies in quantum parameter estimation  have established the existence of  a quantum measurement, mathematically represented by a POVM (Positive-Operator Valued Measure)~\cite{helstrom76,holevo11} that achieves the QCR bound for estimation of a  \emph{single} parameter~\cite{hayashi05,fujiwara06}, while the  quantum bound may not be attainable for  two or more parameters.   Here we propose two measurement schemes which asymptotically attain the QCR bound for both components of the separation over many repetitions. The 
first  is based on the SLIVER scheme of refs.~\cite{nair16,NT16Gaussian} and  approaches the QCR bound for small values of source separation. The second  is a two-dimensional version of the SPADE scheme of ref.~\cite{tsang16a} that in principle attains the bound regardless of the distance between the two sources.
 
This paper is organized as follows. In Sec.~\ref{sec:framework}, we describe the source and system model used in this paper. In Sec.~\ref{sec:bound}, we review the theory of the multiparameter QCR bound and evaluate it to obtain the fundamental limit for the estimation of the Cartesian components of the centroid and separation of the sources. In Sections~\ref{sec:sliver} and \ref{sec:spade}, the SLIVER and SPADE schemes for estimating the components of the separation are detailed, and their Fisher Information (FI) matrices are obtained. In Sec.~\ref{sec:analysis}, we study the performance of the two schemes using Monte-Carlo simulations, and close with concluding remarks in Sec.~\ref{sec:conclusion}.

\section{Source and System Model} \label{sec:framework}
We first lay out the source and imaging system model used in this paper, the former being identical to that in ref.~\cite{tsang16a}. We assume that  two incoherent optical point sources with equal intensities are located on the object plane. Far-field radiation from these sources is collected at the entrance pupil of an optical imaging system such as a microscope or telescope. We assume that the paraxial approximation is valid for the field propagation from object plane to entrance pupil and consider a single polarization only. We further assume that the radiation from the sources is quasi-monochromatic and excites only a single temporal mode in order to focus on the spatial aspects of the resolution problem.  We assume also that the  image-plane coordinates $(x,y)$ have been rescaled by the magnification factor, and that the imaging system is spatially-invariant -- these assumptions entail no essential loss of generality~\cite{goodman05}. Under these conditions, the imaging system is described by its  two-dimensional (possibly complex-valued) point-spread function (PSF)  $\psi(x,y)$, satisying the normalization condition $\int_{-\infty}^\infty\mathrm{d}x \int_{-\infty}^\infty\mathrm{d}y |\psi_s(x,y)|^2=1$ on the image plane. 

We are given two incoherent optical point sources with equal intensities located at coordinates $(X_1, Y_1)$ and $(X_2, Y_2)$ on the object plane. We assume the two sources are such that the probabilities that a single photon emitted by either source arrives at the image plane are equal and given by
\begin{equation}
\epsilon/2 \ll 1.
\end{equation}
We further assume that the probability of more than one photon arriving at the image plane is negligible. 
Under the above assumptions, the quantum density operator of the optical field on the image plane can be written as~\cite{tsang16a}
\begin{equation} \label{eq:rho}
\rho = (1-\epsilon)|\mathrm{vac}\rangle\langle\mathrm{vac}|+\frac{\epsilon}{2}(|\psi_1\rangle\langle\psi_1|+|\psi_2\rangle\langle\psi_2|),
\end{equation}
where $|\mathrm{vac}\rangle$ denotes the vacuum state and the states $\{|\psi_s\rangle\}_{s=1}^2$ are given by
\begin{equation} \label{eq:state_psi}
|\psi_s\rangle = \int_{-\infty}^\infty\mathrm{d}x \int_{-\infty}^\infty\mathrm{d}y \,\psi_s(x,y)|x,y\rangle, \quad s=1,2,
\end{equation}
with the wave functions
\begin{equation} \label{eq:psi}
\psi_s(x,y) = \psi(x-X_s,y-Y_s), \quad s=1,2,
\end{equation}
where $|x,y\rangle$ denotes the state with one photon in the mode corresponding to position $(x,y)$ alone such that $\langle x,y|x',y'\rangle = \delta(x-x')\delta(y-y')$.

Eq.~\eqref{eq:rho} means that a photon arrives with equal probability $\epsilon/2$ from either of the two sources. If a photon arrives from the first source, it is in the state $|\psi_1\rangle$ with wave function $\psi_1(x,y)$; if it comes from the other source,
it is in state $|\psi_2\rangle$ with wave function $\psi_2(x,y)$. The two states are not orthogonal in general, and have the overlap \begin{equation} \label{eq:delta}
\delta \equiv \langle\psi_1|\psi_2\rangle = \int_{-\infty}^\infty\mathrm{d}x \int_{-\infty}^\infty\mathrm{d}y \; \psi_1^*(x,y)\,\psi_2(x,y) \neq 0.
\end{equation}

\begin{figure}
\includegraphics[width=2.058in]{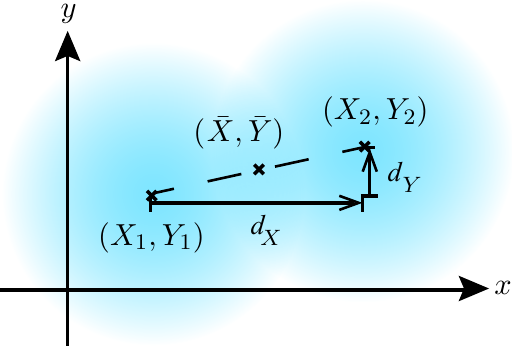}
\caption{\label{fig:parameter}An illustration of the focused image in the image plane of two point sources centered at $(X_1,Y_1)$ and $(X_2,Y_2)$. The shading indicates the approximate extent of the PSF.}
\end{figure}

We also make a realistic simplifying assumption on the PSF $\psi(x,y)$ of the  imaging system, namely that it is symmetric about the origin (or inversion-symmetric), viz.,
\begin{equation} \label{eq:invsym}
\psi(x,y) = \psi(-x,-y),
\end{equation}
for all $x$ and $y$. This assumption is satisfied for most imaging systems of interest, including spatially-invariant systems whose entrance aperture is rectangular or (hard or apodized) circular in shape \cite{goodman05}, and is more general than the assumption of a circularly-symmetric PSF used in ref.~\cite{nair16}. Under this assumption, the overlap $\delta$ of Eq.~\eqref{eq:delta} is real-valued (see Appendix \ref{sec:QFI}).

The parameters we are interested in estimating are the four components of the vector 
\begin{align}
\theta \equiv (\bar X, \bar Y, d_X, d_Y)^\top, \label{eq:theta}
\end{align}
%as the centroid
%\begin{align}
%(\bar X, \bar Y) &= \left(\frac{1}{2}(X_1+X_2), \frac{1}{2}(Y_1+Y_2)\right),
%\intertext{and the separation}
%(d_X, d_Y) &= (X_2-X_1, Y_2-Y_1).
%\end{align}
consisting of the centroid vector
\begin{align}
{(\bar X, \bar Y)} \equiv [{(X_1, Y_1)}+{(X_2, Y_2)}]/2, \label{eq:barXbarY}
\end{align}
and the separation vector
\begin{align}{(d_X, d_Y)} \equiv {(X_2, Y_2)}-{(X_1, Y_1)}, \label{eq:dXdY}
\end{align}
as depicted in Fig. \ref{fig:parameter}.

\section{The quantum limit on two-source localization}  \label{sec:bound}

\subsection{Review of the multiparameter Quantum Cram\'er-Rao (QCR) bound}
Let $\rho_\theta$ be the density operator of a  quantum system depending on an unknown vector parameter  $\theta$. Consider the estimation of $\theta$ from the quantum measurement outcome $\mathcal Y$ on $M$ copies of $\rho_\theta$. The probablity distribution of $\mathcal Y$ is given by
\begin{equation}
\mathrm{P}(\mathcal Y) = \operatorname{tr}[F(\mathcal{Y})\rho_\theta^{\otimes M}],
\end{equation}
where  $F(\mathcal{Y})$ is the positive operator-valued measure (POVM) that characterizes the statistics of the quantum measurement~\cite{helstrom76,wiseman10}. For any estimate $\check\theta(\mathcal{Y})$ of $\theta$ from the measurement outcome, the estimation error covariance matrix has the matrix elements
\begin{align}
\Sigma_{\mu\nu} &\equiv \mathbb E \left\{ \left[\check\theta_\mu(\mathcal Y)-\theta_\mu\right]\left[\check\theta_\nu(\mathcal Y)-\theta_\nu\right]\right\}, 
\end{align}
where $\mathbb E[z(\mathcal Y)]$ of an arbitrary function $z(\mathcal Y)$ of the measurement outcome is the statistical expectation 
\begin{align}
\mathbb E[z(\mathcal Y)]&\equiv \int \mathrm{d}\mathcal Y\, \mathrm P(\mathcal Y) z(\mathcal Y).
\end{align}
%Let the quantum density operator of an object be $\rho$ and is dependent on some unknown parameters $(\theta_1, \theta_2, \dots, \theta_K)$. The measurement outcome $\mathcal Y$ for any quantum measurement on $M$ copies of the object has probability distribution 
%\begin{equation}\mathrm{P}(\mathcal Y) = \operatorname{tr}[F(\mathcal{Y})\rho^{\otimes M}],\end{equation}
%where $\operatorname{tr}$ is the operator trace, $F(\mathcal{Y})$ is the positive operator-valued measure (POVM) that characterizes the measurement and $\rho^{\otimes M}$ denotes the tensor product of $M$ copies of $\rho$. Denotes the estimator of $\theta_\mu$ using $\mathcal Y$ as $\check\theta_\mu(\mathcal Y)$. The estimation error covariance matrix is defined as
%\begin{equation}\Sigma_{\mu\nu} \equiv \sum_\mathcal Y \mathrm P(\mathcal Y) \left[\check\theta_\mu(\mathcal Y)-\theta_\mu\right] \left[\check\theta_\nu(\mathcal Y)-\theta_\nu\right].\end{equation}
For any unbiased estimator, defined as one for which
\begin{equation}
\mathbb{E}\left[\check\theta(\mathcal Y)-\theta\right] = 0,
\end{equation}
 the error covariance matrix $\Sigma$ is bounded by the classical and quantum Cram\'er-Rao bounds~\cite{vantrees01,helstrom76,holevo11,paris09},
\begin{equation}
\Sigma \geq {\mathcal J}^{-1} \geq \mathcal K^{-1}.
\end{equation}
Here, the inequalities mean that matrices $\Sigma - {\mathcal J}^{-1}$, $\Sigma - \mathcal K^{-1}$ and ${\mathcal J}^{-1}-\mathcal K^{-1}$ are positive-semidefinite. Matrix ${\mathcal J}$ is the \emph{classical} Fisher information (FI) matrix given by
\begin{equation} 
{\mathcal J}_{\mu\nu} = \mathbb E \left\{\left[\frac{\partial}{\partial \theta_\mu} \ln \mathrm P(\mathcal Y)\right] \left[\frac{\partial}{\partial \theta_\nu} \ln \mathrm P(\mathcal Y)\right]\right\}, \label{eq:cfi}
\end{equation}
%\begin{equation}j_{\mu\nu} = \sum_\mathcal Y \frac{1}{\mathrm P(\mathcal Y)} \frac{\partial \mathrm P(\mathcal Y)}{\partial \theta_\mu} \frac{\partial \mathrm P(\mathcal Y)}{\partial \theta_\nu}, \label{eq:cfi}\end{equation}
and matrix $\mathcal K$ is the quantum Fisher information (QFI) matrix which can be expressed in terms of the so-called symmetric logarithmic derivative (SLD) operators $\{\mathcal L_\mu\}$ as
\begin{equation}
\mathcal K_{\mu\nu} = M\mathrm{tr}\frac{\mathcal L_\mu \mathcal L_\nu+\mathcal L_\nu \mathcal L_\mu}{2}\rho_\theta. \label{eq:qfi}
\end{equation}
If $\rho_\theta$ is diagonalized in an orthogonal basis $\{|e_n\rangle\}$, viz., $\rho_\theta = \sum_n D_n|e_n\rangle\langle e_n|$, $\mathcal L_\mu$ can be expressed as  \cite{paris09,BC94}
\begin{equation}
\mathcal L_\mu = \sum_{\substack{m,n\\D_m+D_n\neq 0}} \frac{2}{D_m+D_n}\langle e_m|\frac{\partial\rho_\theta}{\partial \theta_\mu}|e_n\rangle |e_m\rangle\langle e_n|. \label{eq:SLD}
\end{equation}

\subsection{Quantum Fisher Information (QFI) Matrix for two-source localization} \label{sec:results_qfi}
We now consider the  problem of estimation of the  centroid  and separation vectors for two incoherent point sources under the model of Sec.~\ref{sec:framework}.  Assuming  the quantum density operator of Eq.~\eqref{eq:rho} and the inversion-symmetry of the PSF (viz., Eq.~\eqref{eq:invsym}), the SLD operators of Eq.~\eqref{eq:SLD} and the QFI matrix $\mathcal K$ of Eq.~\eqref{eq:qfi}  can be explicitly evaluated. The salient details of the derivation of $\mathcal K$ are relegated to  Appendix~\ref{sec:QFI}, namely the basis $\{|e_n\rangle\}$ in which $\rho$ is diagonal, its eigenvalues $\{D_n\}$, and the SLD operators. The QFI matrix in terms of $\theta$ defined in Eq.~\eqref{eq:theta} is found to be
\begin{equation}
\mathcal K = N\begin{pmatrix}4\left(\Delta k_X^2-\gamma_X^2\right) &4\left(\alpha-\gamma_X\gamma_Y\right) &0 &0\\ 4\left(\alpha-\gamma_X\gamma_Y\right) &4\left(\Delta k_Y^2-\gamma_Y^2\right) &0 &0\\ 0 &0 &\Delta k_X^2 &\alpha \\ 0 &0 &\alpha &\Delta k_Y^2\end{pmatrix}, \label{eq:J}
\end{equation}
where $N=M\epsilon$ is the average photon number collected over $M$ trials, and
\begin{align}
\Delta k_X^2 &\equiv \int_{-\infty}^\infty \mathrm dx \int_{-\infty}^\infty \mathrm dy \left|\frac{\partial \psi(x,y)}{\partial x}\right|^2,\nonumber\\
\Delta k_Y^2 &\equiv \int_{-\infty}^\infty \mathrm dx \int_{-\infty}^\infty \mathrm dy \left|\frac{\partial \psi(x,y)}{\partial y}\right|^2,\nonumber\\
\gamma_X &\equiv \int_{-\infty}^\infty \mathrm dx \int_{-\infty}^\infty \mathrm dy\, \psi^*(x-d_X,  y-d_Y)\frac{\partial\psi(x,y)}{\partial x}, \nonumber\\
\gamma_Y &\equiv \int_{-\infty}^\infty \mathrm dx \int_{-\infty}^\infty \mathrm dy\, \psi^*(x-d_X,  y-d_Y)\frac{\partial\psi(x,y)}{\partial y}, \nonumber\\
\alpha &= \operatorname{Re}\left[\int_{-\infty}^\infty \mathrm dx \int_{-\infty}^\infty \mathrm dy \,\frac{\partial \psi^*(x,y)}{\partial x} \frac{\partial \psi(x,y)}{\partial y}\right], \label{eq:parameter}
\end{align}
for $\operatorname{Re}(z)$ denoting the real part of $z$. The quantities $\Delta k_X$ and $\Delta k_Y$ are related to the spatial spectral width of the PSF in the $x$- and $y$-direction respectively and, along with $\alpha$, are independent of the source parameters $\theta$. $\gamma_X$ and $\gamma_Y$ depend on the separation coordinates $(d_X, d_Y)$ but not on the centroid coordinates  ${(\bar X, \bar Y)}$. Thus, $\mathcal K$ as a whole is independent of ${(\bar X, \bar Y)}$, as may be expected from our assumption of a spatially-invariant imaging system. Note that $\mathcal K$ has a block-diagonal form with respect to the  centroid and separation coordinate pairs, and that the  matrix elements related to the estimation errors of separations $d_X$ and $d_Y$---$\mathcal K_{33}$ and $\mathcal K_{44}$---are independent of $(d_X, d_Y)$ as well.

The QFI matrix can be simplified further for the case of a PSF $\psi(x,y)$ with reflection symmetry about $x$- and $y$-axes, viz.,
\begin{align}
\psi(x,y) &= \psi(x,-y) = \psi(-x,-y), \label{eq:psi_reflection}
\end{align}
which is also a sufficient condition for symmetry about the origin (Eq.~\eqref{eq:invsym}).
Under this condition, the quantity $\alpha$ of Eq.~\eqref{eq:parameter} vanishes as its integrand satisfies
\begin{align}
\frac{\partial \psi^*(x,y)}{\partial x} \frac{\partial \psi(x,y)}{\partial y} = -\frac{\partial \psi^*(x,-y)}{\partial x} \frac{\partial \psi(x,-y)}{\partial y}
\end{align}
for all $x,y$; hence, the integral goes to zero. 
The Fisher information matrix $\mathcal K$ becomes
\begin{equation}
\mathcal K = N\begin{pmatrix}4\left(\Delta k_X^2-\gamma_X^2\right) &-4\gamma_X\gamma_Y &0 &0\\ -4\gamma_X\gamma_Y &4\left(\Delta k^2-\gamma_Y^2\right) &0 &0\\ 0 &0 &\Delta k_X^2 &0\\ 0 &0 &0 &\Delta k_Y^2\end{pmatrix}. \label{eq:fisher}
\end{equation}
Note that a circularly-symmetric $\psi(x,y)$ is a sufficient condition for the reflection symmetries. In that case, we have additionally
\begin{equation}
 \Delta k_X = \Delta k_Y  \equiv \Delta k. \label{eq:Delta_k}
\end{equation}
%The quantum Fisher information for separation $d_X$ and $d_Y$ are $\Delta k_X^2$ and $\Delta k_Y^2$

\subsection{Comparison to Direct Imaging}
The QCR bound $\mathcal K^{-1}$ can be compared with the classical CR bound for conventional direct imaging. For direct imaging using an ideal continuum photodetector in the image plane, the probability density of the position of arrival of the photon is expressed in terms of the mean intensity as ~\cite{tsang16a}:
\begin{align}
\Lambda(x,y) &= \frac{1}{2}\big[|\psi_1(x,y)|^2+|\psi_2(x,y)|^2\big], \label{eq:ch5_intensity}
\end{align}
such that the classical FI matrix
\begin{align}
\mathcal J_{\mu\nu}^\textrm{(dir)} &= N \int_{-\infty}^\infty \mathrm{d}x \int_{-\infty}^\infty \mathrm{d}y \frac{1}{\Lambda(x,y)}\frac{\partial \Lambda(x,y)}{\partial \theta_\mu} \frac{\partial \Lambda(x,y)}{\partial \theta_\nu}. \label{eq:ch5_jdirect}
\end{align}
For any PSF $\psi(x,y)$, let
\begin{align}
I(x,y) \equiv |\psi(x,y)|^2
\end{align}
be the intensity point spread function. We assume that the centroid $(\bar X, \bar Y)$ is located at the origin and we are only estimating the separation vector $\eta = (d_X, d_Y)^\top$.
The mean intensity in Eq.~\eqref{eq:ch5_intensity} becomes
\begin{align}
\Lambda(x,y) = \frac{1}{2}\bigg[I\bigg(x+\frac{d_X}{2}, y+\frac{d_Y}{2}) + I(x-\frac{d_X}{2}, y-\frac{d_Y}{2}\bigg)\bigg],
\end{align}
For small values of $d_X$ and $d_Y$, we can expand $\Lambda(x,y)$ to the second order to obtain
\begin{align}
\Lambda(x,y) &= I(x,y) + \frac{d_X^2}{8}\frac{\partial^2 I(x,y)}{\partial x^2}+\frac{d_Xd_Y}{4}\frac{\partial^2 I(x,y)}{\partial x\partial y} \nonumber\\
&\qquad +\frac{d_Y^2}{8}\frac{\partial^2 I(x,y)}{\partial y^2}+o(d^2),
\end{align}
where $o(d^2)$ denotes terms asymptotically smaller than $d_X^2$, $d_Xd_Y$, and $d_Y^2$. 
Substituting this equation into Eq.~\eqref{eq:ch5_jdirect} gives the Fisher information matrix $\mathcal J^\textrm{(dir)}$.
For a circularly symmetric PSF $\psi(x,y)$, the FI matrix in terms of $\eta$ is
\begin{align}
\mathcal J_{11}^\textrm{(dir)} &= \frac{N}{16} (d_X^2\kappa_1+d_Y^2\kappa_2)+o(d^2), \nonumber\\
\mathcal J_{22}^\textrm{(dir)} &= \frac{N}{16} (d_X^2\kappa_2+d_Y^2\kappa_1)+o(d^2), \nonumber\\
\mathcal J_{12}^\textrm{(dir)} &= \frac{N}{16} d_Xd_Y(\kappa_1+\kappa_2)+o(d^2).
\end{align}
where
\begin{align}
\kappa_1 &= \int_{-\infty}^\infty \mathrm{d}x \int_{-\infty}^\infty \mathrm{d}y \frac{1}{I(x,y)} \bigg[\frac{\partial^2 I(x,y)}{\partial x^2}\bigg]^2 \nonumber\\*
&=\int_{-\infty}^\infty \mathrm{d}x \int_{-\infty}^\infty \mathrm{d}y \frac{1}{I(x,y)} \bigg[\frac{\partial^2 I(x,y)}{\partial y^2}\bigg]^2, \nonumber\\
\kappa_2 &= \int_{-\infty}^\infty \mathrm{d}x \int_{-\infty}^\infty \mathrm{d}y \frac{1}{I(x,y)} \bigg[\frac{\partial^2 I(x,y)}{\partial x\partial y}\bigg]^2.
\end{align}

\begin{figure}
\includegraphics[width=3in]{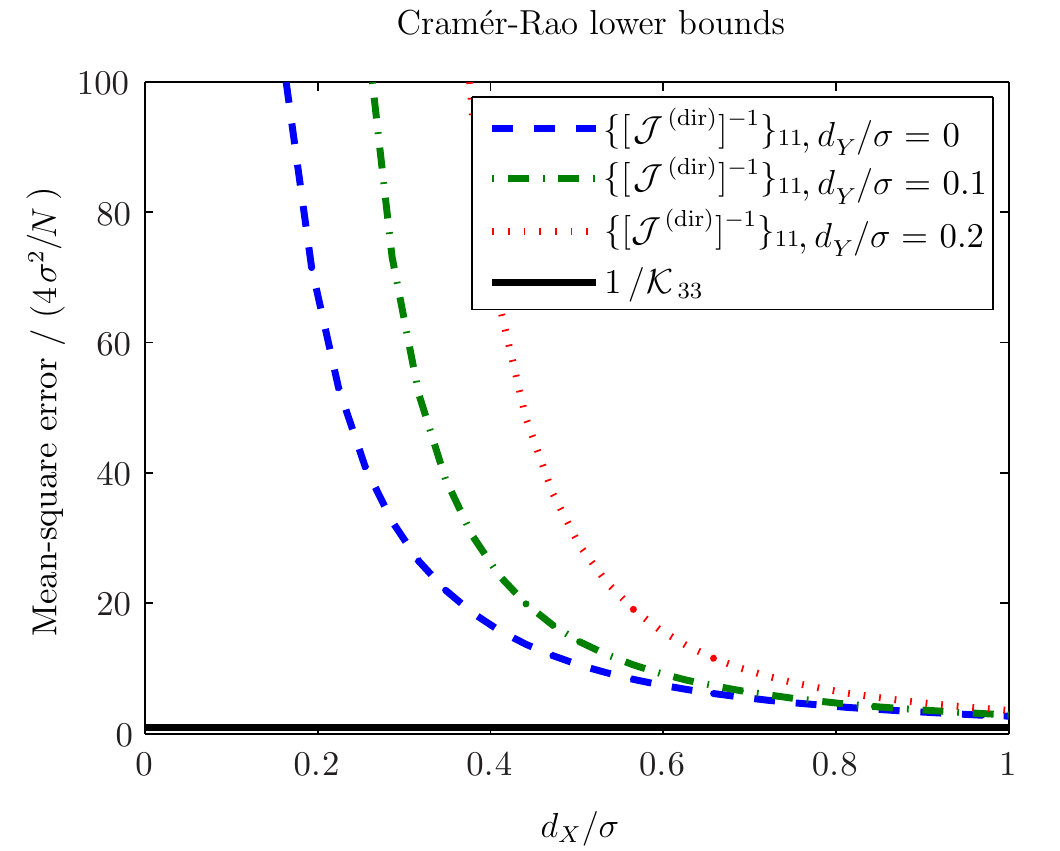}
\caption{\label{fig:direct}Quantum ($1/\mathcal K_{33}$) and classical ($\{[\mathcal J^\textrm{(dir)}]^{-1}\}_{11}$) CR bounds versus normalized separation $d_X/\sigma$ for a circular Gaussian PSD of Eq.~\eqref{eq:directgauss}. The classical bounds are plotted for different value of $d_Y/\sigma =0, 0.1$ and $0.2$. The bounds are normalized with respect to the quantum limit $4\sigma^2/N$.}
\end{figure}

For direct imaging method, the CR bound terms related to the estimation of separation $d_X$ and $d_Y$ are
\begin{align}
\{[\mathcal J^\textrm{(dir)}]^{-1}\}_{11} &\approx \frac{16}{N} \frac{d_X^2\kappa_2+d_Y^2\kappa_1}{(d_X^2-d_Y^2)^2\kappa_1\kappa_2}, \nonumber\\
\{[\mathcal J^\textrm{(dir)}]^{-1}\}_{22} &\approx \frac{16}{N} \frac{d_X^2\kappa_1+d_Y^2\kappa_2}{(d_X^2-d_Y^2)^2\kappa_1\kappa_2}, \label{eq:direct_invj}
\end{align}
which approach infinity as $d_X, d_Y \rightarrow 0$. For illustration, we assume a circular Gaussian PSF $\psi(x,y)$ of the form
\begin{align}
\psi_G(x,y) = \left(\frac{1}{2\pi\sigma^2}\right)^{1/2} \exp\left(-\frac{x^2+y^2}{4\sigma^2}\right), \label{eq:directgauss}
\end{align} 
such that its intensity point spread function
\begin{align}
I_G(x,y) = \frac{1}{2\pi\sigma^2}\exp\left(-\frac{x^2+y^2}{2\sigma^2}\right).
\end{align}
The PSF-dependent terms are now
\begin{align}
\Delta k &= \frac{1}{2\sigma}, \quad
\kappa_1 = 6\kappa_2 = \frac{3}{2\sigma^2},
\end{align}
and Eq.~\eqref{eq:direct_invj} becomes
\begin{align}
\{[\mathcal J^\textrm{(dir)}]^{-1}\}_{11} &\approx \frac{4\sigma^2}{N} \frac{8(d_X/\sigma)^2+48(d_Y/\sigma)^2}{3[(d_X/\sigma)^2-(d_Y/\sigma)^2]^2}, \label{eq:directgauss_invj}\\
\{[\mathcal J^\textrm{(dir)}]^{-1}\}_{22} &\approx \frac{4\sigma^2}{N} \frac{48(d_X/\sigma)^2+8(d_Y/\sigma)^2}{3[(d_X/\sigma)^2-(d_Y/\sigma)^2]^2}. \label{eq:directgauss_invj2}
\end{align}
The QCR bound $1/\mathcal K_{33}$ of Eq.~\eqref{eq:fisher} and the CR bound $\{[\mathcal J^\textrm{(dir)}]^{-1}\}_{11}$ of Eq.~\eqref{eq:directgauss_invj} for the estimation of $d_X$ are plotted as a function of separation parameters $d_X$ and $d_Y$ in Fig.~\ref{fig:direct}. The plot shows a huge divergence of the CR bound for direct imaging method from the quantum limit as $d_X$ decreases. This implies that a considerable improvement can be obtained if a quantum-optimal measurement is implemented. As Eq.~\eqref{eq:directgauss_invj2} is similar to Eq.~\eqref{eq:directgauss_invj} by interchanging the variables $d_X$ and $d_Y$, the plot of the CR bound related to $d_Y$ is identical to Fig.~\ref{fig:direct} and the same conclusion can be drawn for small values of $d_Y$.

In Sec.~\ref{sec:LOschemes}, we discuss concrete measurement schemes to simultaneously estimate  the separation parameters $\eta = (d_X, d_Y)^\top$. For these schemes, we assume that the centroid vector $(\bar X, \bar Y)$ has already been located, and compare their performance to the quantum bound obtained in Sec.~\ref{sec:results_qfi}.

\section{Linear-optics schemes for estimating the separation vector} \label{sec:LOschemes}

In this Section, we give two linear-optics schemes for estimating the separation vector $(d_X,d_Y)$. In so doing, we assume that the centroid of the sources has already been located, perhaps by spatially-resolved direct detection on a portion of the light reaching the image plane -- see, e.g., the hybrid scheme of ref.~\cite{tsang16a}. As is well-known, unlike estimating the separation coordinates, locating the centroid coordinates with direct imaging is near quantum-optimal \cite{helstrom76,tsang16a}.   Assuming that the centroid of the sources is imaged  at the origin of image-plane coordinates, the images of the  sources are centered at $\mp\frac{1}{2}(d_X, d_Y)$ in the image plane. The single-photon wave functions corresponding to the two sources then become
\begin{align} \label{eq:wavefuncs}
\psi_1(x,y) &= \psi\left(x+\frac{d_X}{2}, y+\frac{d_Y}{2}\right), \\
\psi_2(x,y) &=\psi\left(x-\frac{d_X}{2},y-\frac{d_Y}{2}\right).
\end{align}

Our performance analysis of the schemes of this Section -- in particular, the derivation of their classical FI matrices -- proceeds just as well if the two sources have unequal one-photon arrival probabilities $\epsilon_1$ and $\epsilon_2$, which need not necessarily satisfy $\epsilon_1, \epsilon_2 \ll 1$. In this generalization of the source model of Sec.~\ref{sec:framework}, the image-plane density operator of Eq.~\eqref{eq:rho} is replaced by
\begin{align} \label{eq:rhogen}
\rho = (1-\epsilon_1-\epsilon_2)|\mathrm{vac}\rangle\langle\mathrm{vac}|+\epsilon_1|\psi_1\rangle\langle\psi_1|+\epsilon_2|\psi_2\rangle\langle\psi_2|.
\end{align}
We denote the expected number of photons reaching the image plane by $\epsilon_\mathrm{tot} = \epsilon_1+ \epsilon_2$.

\subsection{The two-stage SLIVER scheme} \label{sec:sliver}
We now propose a two-stage interferometric scheme for estimation of $d_X$ and $d_Y$ adapting the SLIVER schemes of refs.~\cite{nair16,NT16Gaussian}. In those works, a thermal source model was adopted  for which the existence of a positive Glauber-Sudarshan $P$ representation allowed an analysis in the framework of semiclassical photodetection theory~\cite{mandel95,Sha09}. The weak single-photon state of Eq.~\eqref{eq:rhogen} does not possess a non-negative $P$ representation, necessitating a  fully quantum analysis that we carry out by propagating the field operators through the system in the Heisenberg picture.

\begin{figure*}
\includegraphics[width=6.7in]{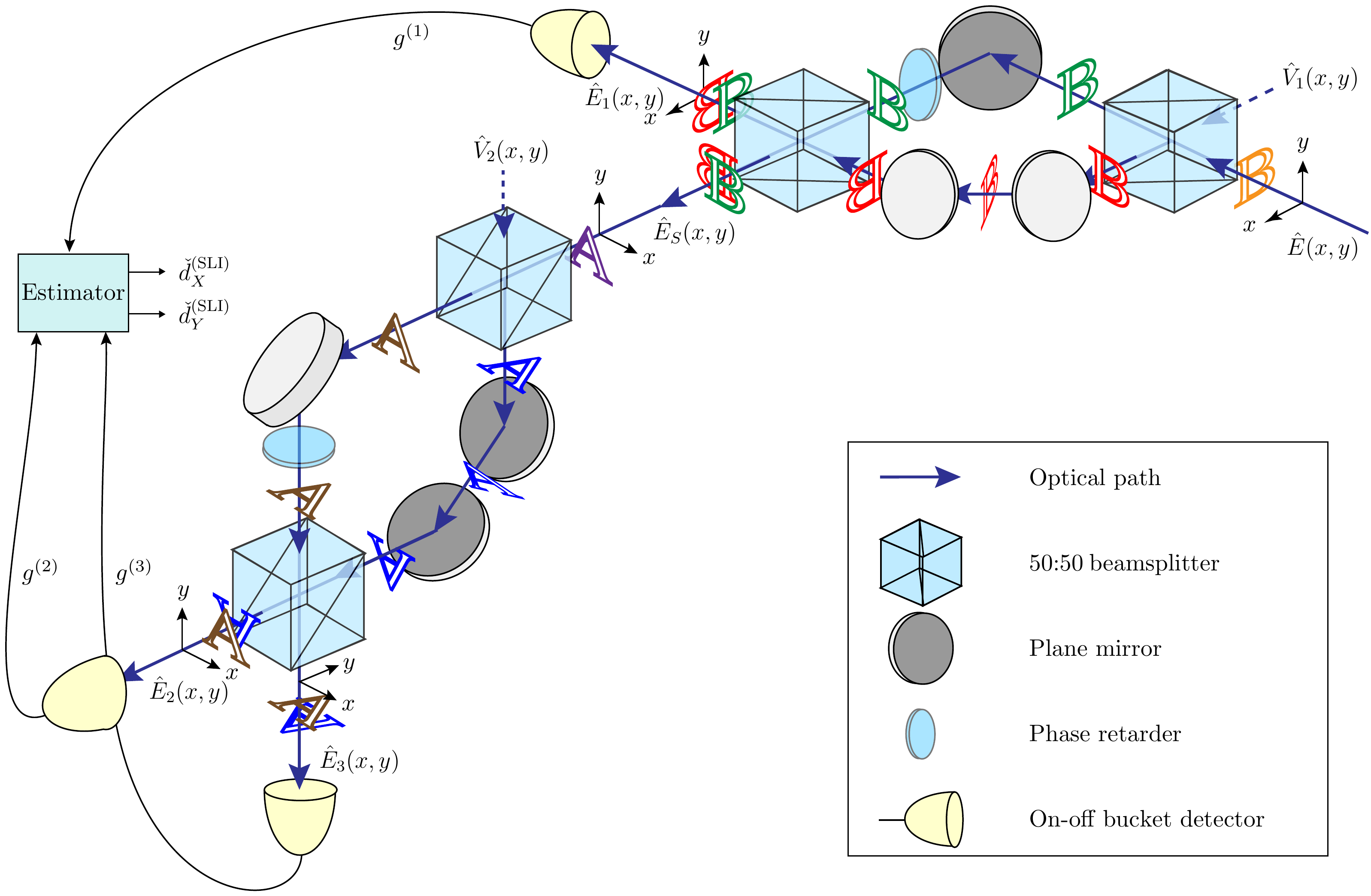}
\caption{\label{fig:SLIVER}A schematic of 2-stage SLIVER -- The image-plane field operator $\hat{E}(x,y)$ is split -- with the appropriate contributions from the field operator $\hat{V}_1(x,y)$ input to the vacuum input port of the first beam splitter --  into its symmetric ($\hat{E}_S(x,y)$) and antisymmetric ($\hat{E}_1(x,y)$) components with respect to reflection about the $y$-axis. The $\hat{E}_1(x,y)$ component impinges upon a bucket photodetector. The $\hat{E}_S(x,y)$ component is separated again into symmetric ($\hat{E}_3(x,y)$) and antisymmetric ($\hat{E}_2(x,y)$) components with respect to reflection about the $x$-axis, which are detected using bucket detectors. The set of binary outcomes, $g^{(1)}$, $g^{(2)}$ and $g^{(3)}$, observed in the detectors over a series of $M$ measurements is processed to give estimates $\check d_X$ and $\check d_Y$ of the components of the separation.  The required field transformations are realized by the extra reflection at an appropriately aligned plane mirror in one arm of the balanced Mach-Zehnder interferometers, which are indicated by the evolution of the letters `$\mathbb{A}$' and `$\mathbb{B}$' through the system.}
\end{figure*}

We assume a PSF $\psi(x,y)$ with reflection symmetries about $x$- and $y$-axes, viz., Eq.~\eqref{eq:psi_reflection}. The SLIVER scheme is  illustrated in Fig.~\ref{fig:SLIVER} and consists of two stages. Viewed semiclassically, the first stage involves the separation of the input field $E(x,y)$ into its antisymmetric component $\left[E(x,y) - E(-x,y)\right]/2$ and its symmetric component $\left[E(x,y) + E(-x,y)\right]/2$  with respect to reflection about the $y$-axis. These components can be obtained by splitting the input field $E(x,y)$ using a 50-50 beamsplitter,  inverting the $x$-coordinates of the field (i.e., reflecting the field about the $y$-axis) in one output and recombining the two beams at a second 50-50 beamsplitter. The optics of this stage thus consists of a balanced Mach-Zehnder interferometer with an extra reflection at an appropriately aligned plane mirror in one arm. 

In the quantum treatment, we replace the complex field amplitudes $E_\alpha(x,y)$ of each beam in any part of the system (indexed generically by $\alpha$) with the corresponding field operators $\hat{E}_\alpha(x,y)$, which are required to satisfy the commutation rules~\cite{mandel95,Sha09}:-
\begin{align} \label{eq:comms}
\left[\hat{E}_\alpha(x,y), \hat{E}_\beta(x',y') \right] &=0, \nonumber \\
\left[\hat{E}_\alpha(x,y), \hat{E}^\dag_\beta(x',y') \right] &= \delta_{\alpha \beta}\; \delta(x-x')\,\delta(y-y').
\end{align}

Using the standard Heisenberg-picture treatment of the input-output relations of a beam splitter~\cite{mandel95,Sha09,GK05quantum}, the output field operators of the first stage of the SLIVER system are given by
\begin{align}
\hat{E}_{1}(x,y) &= \frac{1}{2}\left[\hat{E}(x,y)-\hat{E}(-x,y)\right] \nonumber\\
&\qquad +\frac{1}{2}\left[\hat{V}_1(x,y)+\hat{V}_1(-x,y)\right], \label{eq:E_X} \\
\hat{E}_{S}(x,y) &= \frac{1}{2}\left[\hat{E}(x,y)+\hat{E}(-x,y)\right] \nonumber \\
&\qquad +\frac{1}{2}\left[\hat{V}_1(x,y)-\hat{V}_1(-x,y)\right], \label{eq:E_S}
\end{align}
where $\hat{V}_1(x,y)$ is the input field operator at the open port of the first beamsplitter that must be included to preserve the commutator relations given by Eq.~\eqref{eq:comms} -- the field in this port is in the vacuum state. At the antisymmetric output port with the field operator of Eq.~\eqref{eq:E_X}, an on-off non-spatially-resolving (bucket) detector is placed to distinguish between no photon and one photon. The measurement outcome, denoted $g^{(1)}$, is binary -- zero if the detector does not click and one if it does. 

In the second stage, the output beam $\hat{E}_{S}(x,y)$ of the symmetric port is used as input to a second interferometer which similarly splits the field into antisymmetric ($\hat{E}_{2}(x,y)$) and symmetric ($\hat{E}_{3}(x,y)$) components with respect to reflection about the $x$-axis. The output field operators of the second stage are given by
\begin{align}
\hat{E}_{2}(x,y) &= \frac{1}{2}\left[\hat{E}_{S}(x,y)-\hat{E}_{S}(x,-y)\right] \nonumber\\
&\qquad +\frac{1}{2}\left [\hat{V}_2(x,y)+\hat{V}_2(x,-y)\right], \label{eq:E_Y} \\
\hat{E}_{3}(x,y) &= \frac{1}{2}\left[\hat{E}_{S}(x,y)+\hat{E}_{S}(x,-y)\right] \nonumber \\
&\qquad +\frac{1}{2}\left [\hat{V}_2(x,y)-\hat{V}_2(x,-y)\right], \label{eq:E_R}
\end{align}
where $\hat{V}_2(x,y)$ is the input vacuum field operator at the open port of the first beamsplitter of this stage. The output fields $\hat{E}_{2}(x,y)$ and $\hat{E}_{3}(x,y)$ of this stage impinge upon two on-off bucket detectors to give measurement outcomes $g^{(2)}$ and $g^{(3)}$, respectively. As in the previous stage, $g^{(2)}$ and $g^{(3)}$ take binary values -- 0 if the corresponding detector does not click and 1 if it does -- which are recorded.

The expected photon number at the $r$th detector is given by $\operatorname{tr}(\rho \hat{N}_r)$ where the photon number operator in the $r$-th output beam is given by
\begin{align} \label{eq:Nrdef}
\hat{N}_r &= \int_{-\infty}^\infty \mathrm dx \int_{-\infty}^\infty \mathrm dy \; \hat{E}_r^\dagger(x,y)\hat{E}_r(x,y), \quad r=1,2,3.
\end{align}
Since the state $\rho$ of Eq.~\eqref{eq:rhogen} has at most one photon, at most one photon will impinge upon the three photodetectors taken together.  Therefore, there are only four mutually exclusive measurement outcomes  -- outcome `0' corresponds to the case where no photon detected in any of the three detectors and outcome `$r$' to the case where only the $r$-th detector clicks.  The probabilities of these outcomes are
\begin{align}
P(0) &= \operatorname{Pr}[g^{(1)} = 0, g^{(2)} = 0, g^{(3)} = 0], \nonumber\\
P(1) &= \operatorname{Pr}[g^{(1)} = 1, g^{(2)} = 0, g^{(3)} = 0], \nonumber\\
P(2) &= \operatorname{Pr}[g^{(1)} = 0, g^{(2)} = 1, g^{(3)} = 0], \nonumber\\
P(3) &= \operatorname{Pr}[g^{(1)} = 0, g^{(2)} = 0, g^{(3)} = 1].
\end{align}
Since either zero or one photon arrives at each detector, the probability that the $r$-th detector clicks is equal to the expected photon number at the $r$th detector, i.e.,
\begin{align} \label{eq:Pr}
P(r) &= \operatorname{tr}(\rho \hat{N}_r), \;\; r=1,2,3, \\
P(0) &= 1 -\epsilon_\mathrm{tot}.
\end{align} 
The calculation of the above probabilities is detailed in Appendix~\ref{sec:SLIVER}, with the result:
\begin{align}
P(1) &= \frac{\epsilon_\mathrm{tot}}{2}(1-\delta_x), \nonumber\\
P(2) &= \frac{\epsilon_\mathrm{tot}}{4}(1+\delta_x-\delta_y-\delta),  \label{eq:Pr_sol}\\
P(3) &= \frac{\epsilon_\mathrm{tot}}{4}(1+\delta_x+\delta_y+\delta),\nonumber
\end{align}
where
\begin{align}
\delta_x &\equiv \int_{-\infty}^\infty \mathrm{d}x \int_{-\infty}^\infty \mathrm{d}y \,\psi^*(x,y)\psi(x-d_X,y), \label{eq:deltaX} \\
\delta_y &\equiv \int_{-\infty}^\infty \mathrm{d}x \int_{-\infty}^\infty \mathrm{d}y \,\psi^*(x,y)\psi(x,y-d_Y), \label{eq:deltaY}
\end{align}
and $\delta$ is defined in Eq.~\eqref{eq:delta}.

The classical FI matrix $\mathcal J^\mathrm{(SLI)}$  for the separation vector $\eta = (d_X, d_Y)^\top$ using  SLIVER has the elements (cf. Eq.~\eqref{eq:cfi}):
\begin{align} \label{eq:j_SLI}
{\mathcal J}_{\mu\nu}^\mathrm{(SLI)} &= \sum_{r=1}^3 P(r)\frac{\partial \ln P(r)}{\partial \eta_\mu}\frac{\partial \ln P(r)}{\partial \eta_\nu}, \quad \mu,\nu = 1,2.
\end{align}
Using Eqs.~\eqref{eq:j_SLI} and \eqref{eq:Pr_sol}, we have
\begin{widetext}
\begin{align} \label{eq:jelements_SLI}
{\mathcal J}_{11}^\mathrm{(SLI)} &= \frac{\epsilon_\mathrm{tot}}{2}\left[\frac{1}{1-\delta_x}\left(\frac{\partial \delta_x}{\partial d_X}\right)^2 +\frac{1}{2(1+\delta_x-\delta_y-\delta)}\left(\frac{\partial \delta_x}{\partial d_X}-\frac{\partial \delta}{\partial d_X}\right)^2+\frac{1}{2(1+\delta_x+\delta_y+\delta)}\left(\frac{\partial \delta_x}{\partial d_X}+\frac{\partial \delta}{\partial d_X}\right)^2\right], \nonumber\\
{\mathcal J}_{22}^\mathrm{(SLI)} &= \frac{\epsilon_\mathrm{tot}}{4}\left[\frac{1}{1+\delta_x-\delta_y-\delta}\left(\frac{\partial \delta_y}{\partial d_Y}+\frac{\partial \delta}{\partial d_Y}\right)^2+\frac{1}{1+\delta_x+\delta_y+\delta}\left(\frac{\partial \delta_y}{\partial d_Y}+\frac{\partial \delta}{\partial d_Y}\right)^2\right], \nonumber\\
{\mathcal J}_{12}^\mathrm{(SLI)} 
&= \frac{\epsilon_\mathrm{tot}}{4}\left[\frac{1}{1+\delta_x+\delta_y+\delta}\left(\frac{\partial \delta_x}{\partial d_X}+\frac{\partial \delta}{\partial d_X}\right)\left(\frac{\partial \delta_y}{\partial d_Y}+\frac{\partial \delta}{\partial d_Y}\right)-\frac{1}{1+\delta_x-\delta_y-\delta}\left(\frac{\partial \delta_x}{\partial d_X}-\frac{\partial \delta}{\partial d_X}\right)\left(\frac{\partial \delta_y}{\partial d_Y}+\frac{\partial \delta}{\partial d_Y}\right)\right] \nonumber\\
&= {\mathcal J}_{21}^\mathrm{(SLI)}.
\end{align}
\end{widetext}
%where $\delta_x$ is given by Eq.~\eqref{eq:deltaX}$, \delta_y$ is given by Eq.~\eqref{eq:deltaY} and $\delta$ is given by Eq.~\eqref{eq:delta}. 

%The expression for  $j^\mathrm{(SLI)}$ itself is too complex to be included here. However, as $d_X, d_Y \rightarrow 0$, we show in  Appendix~\ref{sec:SLIVER} that  the matrix elements approach
%\begin{align}
%j_{11}^\mathrm{(SLI)} &\rightarrow \epsilon_\mathrm{tot}\Delta k^2 = \mathcal K_{33}, \nonumber\\
%j_{22}^\mathrm{(SLI)} &\rightarrow \epsilon_\mathrm{tot}\Delta k^2 = \mathcal K_{44},  \nonumber\\
%j_{12}^\mathrm{(SLI)} &= j_{21}^\mathrm{(SLI)} \rightarrow 0,
%\end{align}
%where $\Delta k$ is given by Eq.~\eqref{eq:Delta_k}, and $\mathcal K_{33}$ and $\mathcal K_{44}$ are the QFI matrix elements  in Eq.~\eqref{eq:J} related to the separation parameters. Thus, $j^\mathrm{(SLI)}$ approaches the QCR bound  $\mathcal K$ for small $d_X$ and $d_Y$, and the extended SLIVER scheme approaches the two-parameter error bound given by quantum theory. 

\begin{figure}
\includegraphics[width=3in]{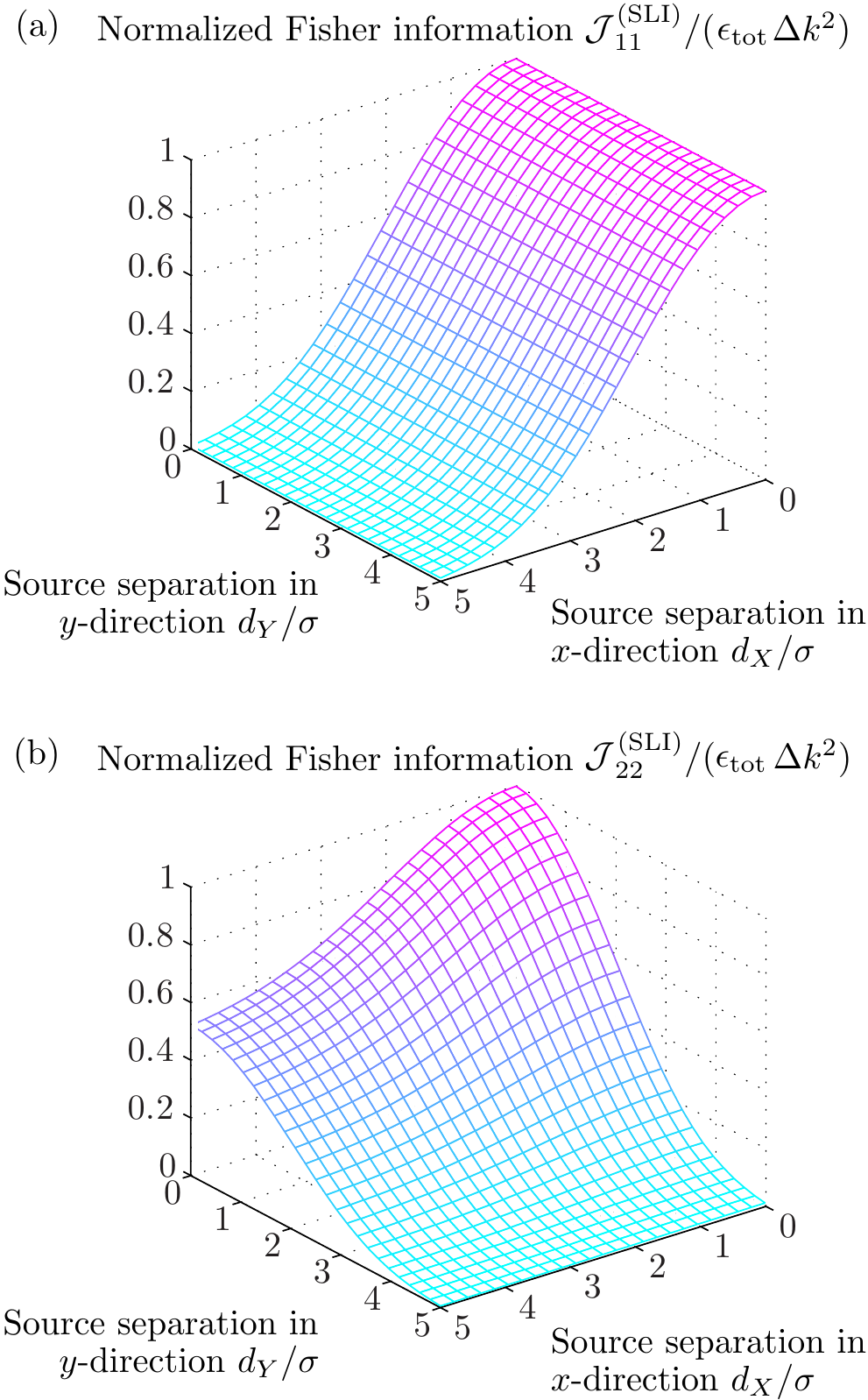}
\caption{\label{fig:Fisher} The (classical) Fisher information matrix $\mathcal J^{\mathrm{(SLI)}}$ for the SLIVER scheme as a function of the source separation  $(d_X ,d_Y)$. (a)  Fisher information for $x$-separation ${\mathcal J}_{11}^\mathrm{(SLI)}$. (b)  Fisher information for $y$-separation ${\mathcal J}_{22}^\mathrm{(SLI)}$. The plots are normalized with respect to the value $\epsilon_\mathrm{tot}\Delta k^2$ of $\mathcal K_{33}$ and $\mathcal K_{44}$. The quantum bound is attained at $d_X=d_Y=0$ as illustrated in (a) and (b). The circular Gaussian PSF of Eq. \eqref{eq:circGauss} is assumed, with the total source strength $\epsilon_\mathrm{tot} = 2\times 10^{-3}$ photons, and the plots are independent of the half-width $\sigma$.}
\end{figure}

We illustrate the above results for a circular Gaussian PSF:-
\begin{equation} \label{eq:circGauss}
\psi_G(x,y) = \left(\frac{1}{2\pi\sigma^2}\right)^{1/2} \exp\left(-\frac{x^2+y^2}{4\sigma^2}\right).
\end{equation}
The PSF-dependent quantities appearing in the FI matrix are then given by
{\allowdisplaybreaks\begin{align}
\delta &= \delta_x\delta_y, \nonumber\\
\delta_x &=\exp\left(-\frac{d_X^2}{8\sigma^2}\right), \quad
\delta_y =\exp\left(-\frac{d_Y^2}{8\sigma^2}\right), \nonumber\\
\frac{\partial \delta_x}{\partial d_X} &= -\frac{d_X}{4\sigma^2}\exp\left(-\frac{d_X^2}{8\sigma^2}\right), \nonumber\\
\frac{\partial \delta_y}{\partial d_Y} &= -\frac{d_Y}{4\sigma^2}\exp\left(-\frac{d_Y^2}{8\sigma^2}\right), \nonumber\\
\frac{\partial \delta}{\partial d_X} &= \delta_y\frac{\partial \delta_x}{\partial d_X}, \qquad \frac{\partial \delta}{\partial d_Y} = \delta_x\frac{\partial \delta_y}{\partial d_Y}.
\end{align}
}In this example, the FI matrix $\mathcal J^\mathrm{(SLI)}$ has elements
\begin{align}\label{eq:JcircGauss}
{\mathcal J}_{11}^\mathrm{(SLI)} &= \frac{\epsilon_\mathrm{tot}}{1-\delta_x^2}\left(\frac{\partial \delta_x}{\partial d_X}\right)^2, \nonumber\\
{\mathcal J}_{22}^\mathrm{(SLI)} &= \frac{\epsilon_\mathrm{tot}}{1-\delta_y^2}\frac{1+\delta_x}{2}\left(\frac{\partial \delta_y}{\partial d_Y}\right)^2, \nonumber\\
{\mathcal J}_{12}^\mathrm{(SLI)} &= {\mathcal J}_{21}^\mathrm{(SLI)} =0.
\end{align}
As $d_X, d_Y \rightarrow 0$, matrix elements approach
\begin{align} \label{eq:limitsliver}
{\mathcal J}_{11}^\mathrm{(SLI)} &\rightarrow \epsilon_\mathrm{tot}\Delta k^2 = \mathcal K_{33}, \nonumber\\
{\mathcal J}_{22}^\mathrm{(SLI)} &\rightarrow \epsilon_\mathrm{tot}\Delta k^2 = \mathcal K_{44}, 
\end{align}
where $\Delta k^2 = (4\sigma^2)^{-1}.$
The FI elements ${\mathcal J}_{11}^\mathrm{(SLI)}$ and  ${\mathcal J}_{22}^\mathrm{(SLI)}$ of Eq.~\eqref{eq:JcircGauss} are plotted as a function of separation parameters $d_X$ and $d_Y$ in Fig.~\ref{fig:Fisher}. The total source strength $\epsilon_\mathrm{tot} = 2\times 10^{-3}$ photons. The plots are normalized to $\epsilon_\mathrm{tot}\Delta k^2$, the values of $\mathcal K_{33}$ and $\mathcal K_{44}$. We see that the maximum values of ${\mathcal J}_{11}^\mathrm{(SLI)}$ and ${\mathcal J}_{22}^\mathrm{(SLI)}$, attained at $d_X=d_Y=0$, are equal to the value of QCR bound  obtained in Sec.~\ref{sec:bound} for the case of $\epsilon_1 = \epsilon_2$.

Eqs.~\eqref{eq:JcircGauss} indicate, as seen in Fig.~\ref{fig:Fisher}(a) and Fig.~\ref{fig:Fisher}(b),  that the FI on $d_X$ -- ${\mathcal J}_{11}^\mathrm{(SLI)}$ -- remains unchanged despite variation in $d_Y$, while the FI on $d_Y$ -- ${\mathcal J}_{22}^\mathrm{(SLI)}$ -- depends on values of both $d_X$ and $d_Y$. This asymmetry is a consequence of our estimating $d_X$ in the first stage of the scheme and $d_Y$ in the second. 

A  simpler non-cascaded version of the  scheme may be envisaged in which the input field $\hat{E}(x,y)$ is split using a 50:50 beamsplitter, and the two outputs are used to separately estimate $d_X$ and $d_Y$. Though  it treats the separation components symmetrically, such a setup can only approach half of the QCR bound for each component due to the energy splitting. On the other hand, in the cascaded scheme given here, if $d_X \approx 0$, $\delta_x \approx 1$, so that $P(1) \approx 0$, and the single photon, when present in the input field, is available with high probability for estimating $d_Y$  in the second stage, allowing the composite setup to  approach the QCR bound for sub-Rayleigh separations.

\subsection{Spatial-mode demultiplexing (SPADE)}\label{sec:spade}
We now generalize the SPADE scheme of ref.~\cite{tsang16a} to the estimation of the vector separation. We assume the circular Gaussian PSF
\begin{align} \label{eq:GPSF}
\psi_G(x,y) = \left(\frac{1}{2\pi\sigma^2}\right)^{1/2} \exp\left(-\frac{x^2+y^2}{4\sigma^2}\right).
\end{align}
In the derivation of the QFI matrix $\mathcal K$ in Sec.~\ref{sec:bound}, we worked in the orthonormal basis given by Eq.~\eqref{eq:basis}. Now consider the discrete Hermite-Gaussian (HG) basis $\{|\phi_{qr}\rangle; {q,r=0,1,\dots}\}$ of wave functions for the one-photon subspace, where
\begin{align}
|\phi_{qr}\rangle &=\int_{-\infty}^\infty\mathrm{d}x \int_{-\infty}^\infty\mathrm{d}y \,\phi_{qr}(x,y)|x,y\rangle, \\
\phi_{qr}(x,y) &= \left(\frac{1}{2\pi\sigma_X\sigma_Y}\right)^{1/2} \frac{1}{\sqrt{2^{q+r} q!r!}} H_q\left(\frac{x}{\sqrt{2}\sigma_X}\right)\nonumber\\
&\quad \times H_r\left(\frac{y}{\sqrt{2}\sigma_Y}\right) \exp\left(-\frac{x^2}{4\sigma_X^2}-\frac{y^2}{4\sigma_Y^2}\right), \label{eq:HermiteGauss}
\end{align}
where $H_q$ and $H_r$ are the Hermite polynomials~\cite{yariv89} for $q,r = 0,1,\dots$, $\sigma_X = \sigma \neq \sigma_Y \equiv s\sigma$. The significance of $s \neq 1$ will appear shortly. 
Since the Hermite-Gaussian functions are an orthonormal basis for the space of wave functions in the image plane, the   projections
\begin{align} 
W_0 &= |\mathrm{vac}\rangle\langle\mathrm{vac}|, \nonumber \\
 \quad W_1(q,r) &= |\phi_{qr}\rangle\langle\phi_{qr}|,\;\; q,r = 0, 1,\ldots, \label{eq:POVM}
\end{align}
together form a POVM on the vacuum$+$one-photon subspace of the image-plane field.

The transformation 
\begin{align}
E(x,y) \mapsto E'(x,y) = s^{-1/2}E(x,y/s)
\end{align}
 on the space of image-plane wave functions is unitary, and takes the PSF of Eq.~\eqref{eq:GPSF} to the elliptical Gaussian
\begin{align} 
\psi'(x,y) &= \left(\frac{1}{2\pi\sigma_X\sigma_Y}\right)^{1/2}\exp\left(-\frac{x^2}{4\sigma_X^2}-\frac{y^2}{4\sigma_Y^2}\right) \label{eq:psiprime} \\
&= \phi_{00}(x,y).
\end{align}
It also induces a unitary transformation $\hat{U}_s$ on the one-photon subspace transforming the state of Eq.~\eqref{eq:rhogen} to
\begin{align}
\rho' &= \hat{U}_s \rho \,\hat{U}_s^\dag,\\
&=  (1-\epsilon_\mathrm{tot})|\mathrm{vac}\rangle\langle\mathrm{vac}|+\epsilon_1|\psi'_1\rangle\langle \psi'_1|+\epsilon_2|\psi'_2\rangle\langle\psi'_2|,
\end{align} 
where, from  Eq.~\eqref{eq:wavefuncs},
\begin{align}
\psi'_1(x,y) &= \psi'\left(x+\frac{d_X}{2}, y+\frac{s\,d_Y}{2}\right), \\
\psi'_2(x,y) &= \psi'\left(x-\frac{d_X}{2}, y-\frac{s\,d_Y}{2}\right).
\end{align}
The POVM \eqref{eq:POVM}, if  performed on the state $\rho'$, has the outcome probabilities 
\begin{align}
P_0 &\equiv \mathrm{tr}(W_0\rho') = 1-\epsilon_\mathrm{tot}, \\
P_1(q,r) &\equiv \mathrm{tr}[W_1(q,r)\rho'] \\ &= \epsilon_1\,|\langle\phi_{qr}|\psi'_1\rangle|^2+ \epsilon_2 \,|\langle\phi_{qr}|\psi'_2\rangle|^2. \label{eq:overlaps}
\end{align}
The overlaps in Eq. \eqref{eq:overlaps} 
\begin{align}
|\langle\phi_{qr}|\psi'_1\rangle|^2 &= \left| \int_{-\infty}^\infty\mathrm{d}x \int_{-\infty}^\infty\mathrm{d}y \,\phi_{qr}^*(x,y) \right.\nonumber\\
&\qquad \times\left. \phi_{00}\left(x+\frac{d_X}{2}, y+\frac{sd_Y}{2}\right) \right|^2,\\
|\langle\phi_{qr}|\psi'_2\rangle|^2 &= \left| \int_{-\infty}^\infty\mathrm{d}x \int_{-\infty}^\infty\mathrm{d}y \,\phi_{qr}^*(x,y) \right.\nonumber\\
&\qquad \times\left. \phi_{00}\left(x-\frac{d_X}{2}, y-\frac{sd_Y}{2}\right) \right|^2.
\end{align}
 can be evaluated as in ref.~\cite{tsang16a} using properties of Hermite polynomials, viz.,
\begin{align}
|\langle\phi_{qr}|\psi'_1\rangle|^2 &= |\langle\phi_{qr}|\psi'_2\rangle|^2= \exp(-Q-R)\frac{Q^q R^r}{q! r!},
\intertext{where}
Q &= \frac{d_X^2}{16\sigma^2}, \qquad R = \frac{d_Y^2}{16\sigma^2},
\end{align}
so that the probability
\begin{equation} \label{eq:spadeprob}
P_1(q,r) = \epsilon_\mathrm{tot} \exp(-Q-R)\frac{Q^q R^r}{q! r!}.
\end{equation}
Using Eq.~\eqref{eq:cfi}, the FI matrix for the HG-basis measurement on $\eta = (d_X, d_Y)^\top$ can be calculated. Its matrix elements are
\begin{align}
{\mathcal J}_{11}^\mathrm{(HG)} &= \sum_{q,r = 0}^\infty P_1(q,r)\left[\frac{\partial}{\partial d_X} \ln P_1(q,r) \right]^2 \\
&=\frac{\epsilon_\mathrm{tot}}{Q}\left(\frac{\partial Q}{\partial d_X}\right)^2 = \frac{\epsilon_\mathrm{tot}}{4\sigma^2} =\mathcal K_{33}, \\
{\mathcal J}_{22}^\mathrm{(HG)} &= \sum_{q,r = 0}^\infty P_1(q,r)\left[\frac{\partial}{\partial d_Y} \ln P_1(q,r) \right]^2 \\
&=\frac{\epsilon_\mathrm{tot}}{R}\left(\frac{\partial R}{\partial d_Y}\right)^2 = \frac{\epsilon_\mathrm{tot}}{4\sigma^2} = \mathcal K_{44}, \\
{\mathcal J}_{12}^\mathrm{(HG)} &= {\mathcal J}_{21}^\mathrm{(HG)} =0,
\end{align}
which exactly equals the QFI matrix given by Eq.~\eqref{eq:fisher} if $\epsilon_1=\epsilon_2$. This proves that the POVM \eqref{eq:POVM} is optimal for a Gaussian PSF. 

\begin{figure}
\includegraphics[width=3.5in]{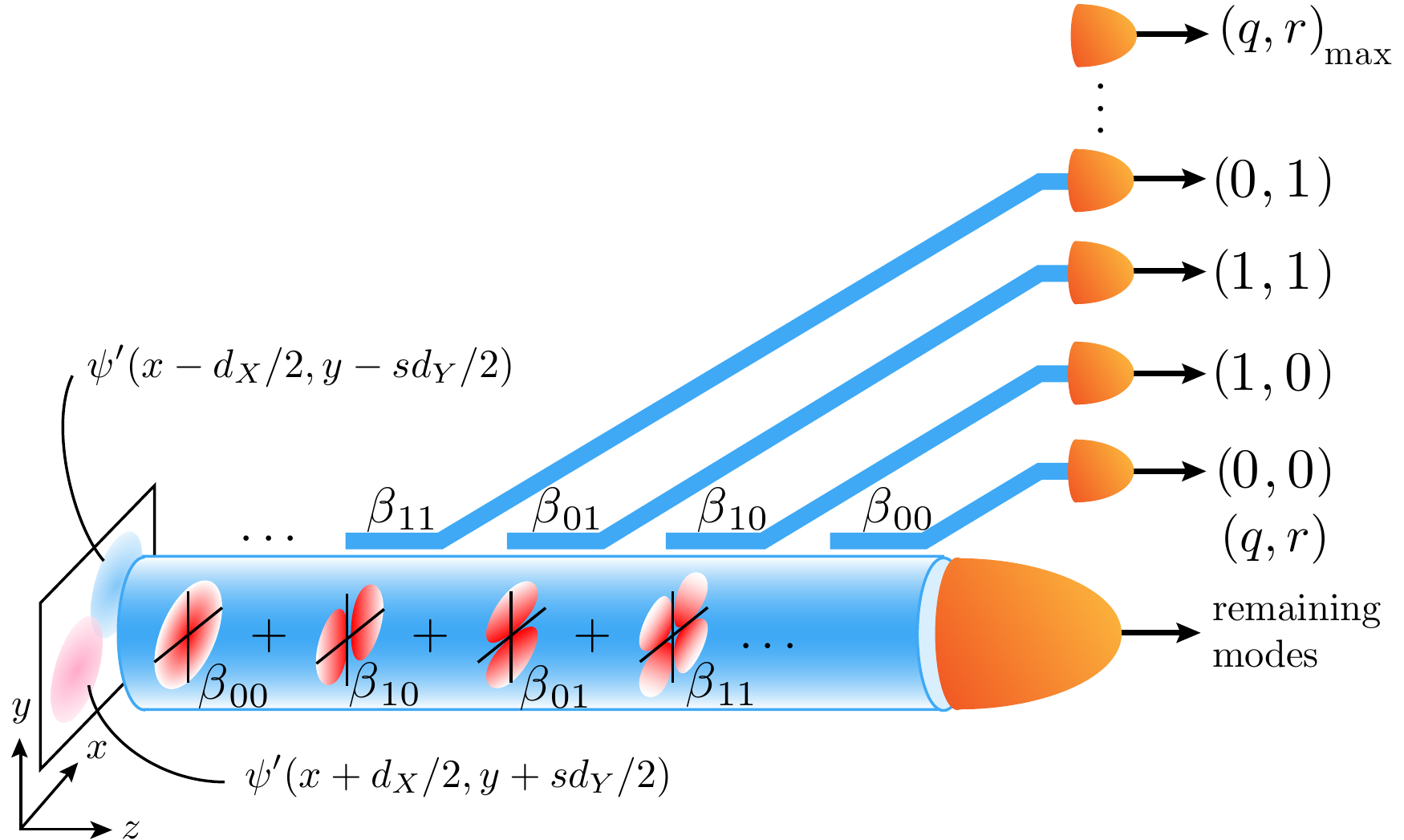}
\caption{\label{fig:spade} A schematic drawing of a fiber-optic implementation of SPADE. The  image-plane field, scaled in the $y$-direction  is coupled into an elliptical multimode fiber with nondegenerate propagation constants for each of the HG basis modes $\phi_{qr}(x,y)$. Using evanescent coupling, each mode is coupled to an individual single-mode waveguide of specific propagation constant terminated by an on-off detecor. The photon counter at the end of the multimode fiber captures any remaining photon in the higher-order or leaky modes.}
\end{figure}

It remains to show how to implement the POVM \eqref{eq:POVM} corresponding to SPADE using linear optics. The image-plane field in Eq.~\eqref{eq:rho} is first scaled in one direction with a series of  mirrors such that the PSF $\psi'(x,y)$ of this augmented system is of the form of Eq.~\eqref{eq:psiprime}.
A quadratic-index fiber can support the Hermite-Gaussian mode profiles~\cite{zhang13}. A cylindrical fiber will have degnerate propagation constants $\beta_{qr}$ for modes with the same total order $(q+r)$. Therefore, the optical field is coupled into an \emph{elliptical} multi-mode fiber supporting the modes in Eq.~\eqref{eq:HermiteGauss}. If  the scaling factor $s$ is chosen carefully, each mode will have a distinct propagation constant $\beta_{qr}$ along the propagation direction $z$. The field in the elliptical fiber is then coupled to different single-mode waveguides with matching propagation constants via evanescent coupling as illustrated in Fig.~\ref{fig:spade}. The phase matching condition ensures that only one mode from the elliptical fiber is coupled to each waveguide, which are then detected using  individual on-off detectors in the far-field.  In theory, $s$ needs to be an irrational number, but to break the degeneracy for a large enough number of modes, $s$ can be taken to be a rational number with a large denominator.

\section{Monte-Carlo analysis of SLIVER and SPADE} \label{sec:analysis}
To demonstrate that the two schemes of Sec.~\ref{sec:LOschemes}  perform as predicted by their CR bounds, we implement Monte-Carlo simulations of the mean-square error (MSE) for SLIVER and SPADE using maximum likelihood (ML) estimators~\cite{vantrees01}. In a sequence of $M$ measurements, the shots in which no photon arrives (which happens with probability $(1-\epsilon_\mathrm{tot})$ in each shot) are uninformative and can only be discarded. Therefore, it is convenient to condition our analysis on a fixed number $L$ of detected photons. In doing so, instead of considering $M$ copies of the state $\rho$ of Eq.~\eqref{eq:rho}, we are effectively considering $L$ copies of the conditional single-photon state
\begin{align}
\rho' = \frac{1}{2}(|\psi_1\rangle\langle\psi_1|+|\psi_2\rangle\langle\psi_2|),
\end{align}
where we have assumed for concreteness that the sources are equally strong. It is readily verified that the (per-shot) conditional QFI and FI matrices for SLIVER and SPADE are obtained by simply dividing the unconditional ones calculated previously by $\epsilon$. In particular, the QFI matrix  for estimation of $d_X$ and $d_Y$ using $L$ copies of $\rho'$  becomes $\mathrm{diag}(L/4\sigma^2, L/4\sigma^2)$ so that the QCR  bound for each separation parameter is $4\sigma^2/L$. 

In the following, we will adopt this approach for analyzing the performance of SLIVER and SPADE.
For all the simulations, the circular Gaussian PSF of Eq.~\eqref{eq:circGauss} is assumed, and each MSE is computed by averaging over $10^5$ Monte-Carlo runs.

\subsection{Monte-Carlo analysis of SLIVER}
In $M$ trials, consider direct detection of $\hat{E}_1(x,y)$, $E_2(x,y)$ and $E_3(x,y)$ using three on-off bucket detectors as in Fig.~\ref{fig:SLIVER}. Suppose $L$ trials result in photon detections and are postselected, and indexed by $l \in \{1,\ldots,,L\}$. The postselected measurement record consists of the bitstrings $(g_1^{(1)}, g_2^{(1)}, \dots, g_L^{(1)})$, $(g_1^{(2)}, g_2^{(2)}, \dots, g_L^{(2)})$ and $(g_1^{(3)}, g_2^{(3)}, \dots, g_L^{(3)})$, where $g_l^{(1)}$, $g_l^{(2)}$ and $g_l^{(3)}$ are zero (one) if the corresponding detector did not click (clicked) in the $l$-th postselected trial. 

The total numbers of clicks observed in the three detectors  are respectively
\begin{align}
G^{(1)} &= \sum_{l=1}^L g_l^{(1)}, \quad
G^{(2)} = \sum_{l=1}^L g_l^{(2)}, \quad
G^{(3)} = \sum_{l=1}^L g_l^{(3)},
\end{align}
with $L = G^{(1)} +G^{(2)} +G^{(3)}$. For the circular Gaussian PSF, the ML estimators for $d_X$ and $d_Y$ can be shown to be
\begin{align} \label{eq:ML_SLI}
\check d_{X}^\mathrm{(SLI)} &= \begin{cases} 2\sigma \sqrt{-2\ln \left(1-\frac{2G^{(1)}}{L}\right)} &\quad \text{if } \frac{2G^{(1)}}{L} < 1, \\
2\sigma &\quad \text{otherwise},\end{cases} \nonumber\\
\check d_{Y}^\mathrm{(SLI)} &= \begin{cases} 2\sigma \sqrt{-2\ln \left(1-\frac{2G^{(2)}}{L-G^{(1)}}\right)} &\quad \text{if } \frac{2G^{(2)}}{L-G^{(1)}} < 1, \\
2\sigma &\quad \text{otherwise}.\end{cases}
\end{align}
The second case for both $\check d_X^\mathrm{(SLI)}$ and $\check d_Y^\mathrm{(SLI)}$ is necessary because the logarithm function $\ln(z)$ in the equations for the estimators is undefined for $z \leq 0$. The estimators are set to an arbitrary value in that event, which happens with vanishing probability as $L$ increases.

\begin{figure}
\includegraphics[width=2.5in]{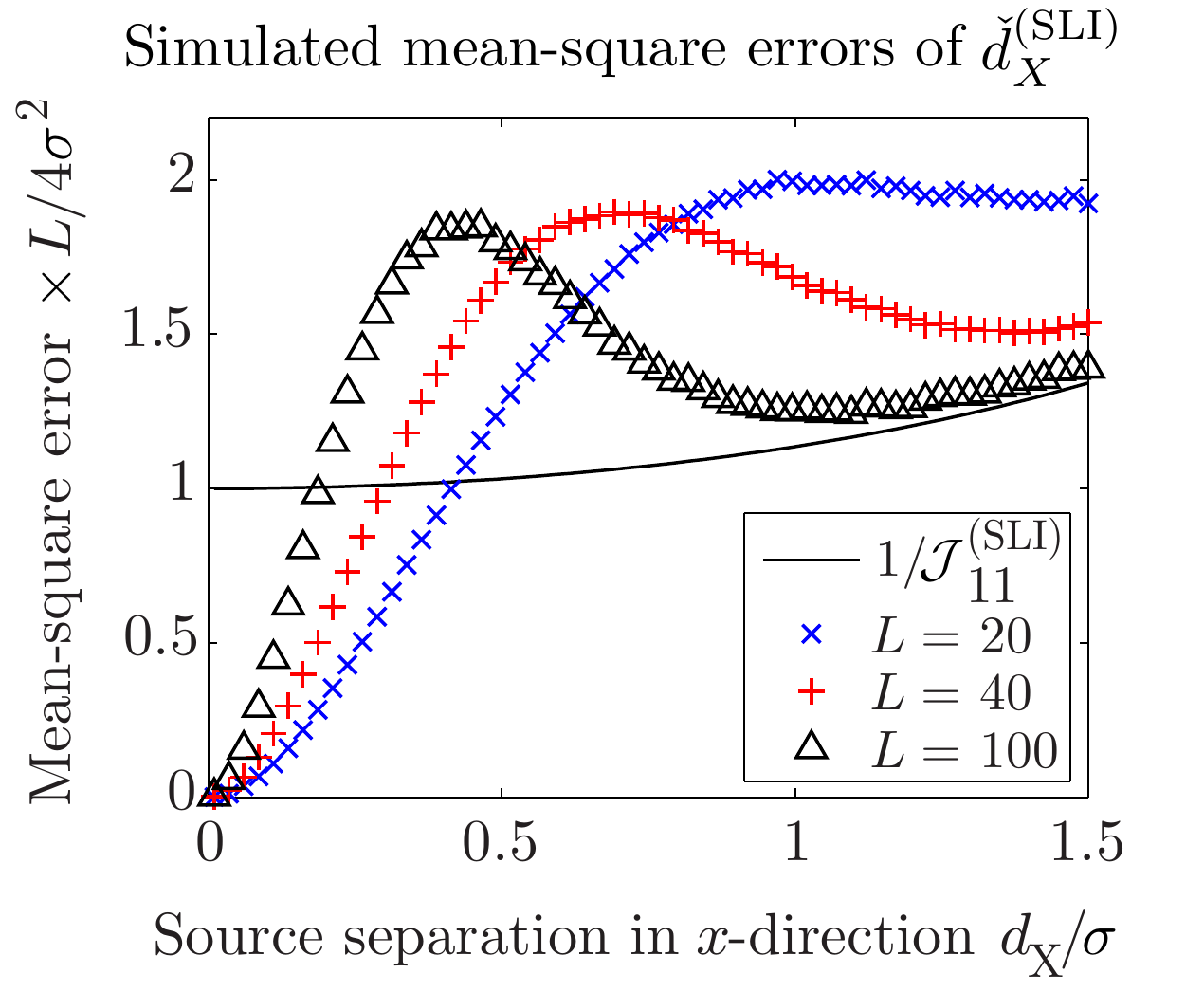}
\caption{\label{fig:MSE_SLI}Simulated mean-square errors of SLIVER with maximum likelihood estimator of  Eq.~\eqref{eq:ML_SLI}. The MSE of estimator $\check d_X^\mathrm{(SLI)}$ as a function of the separation in $x$-direction for $L = 20, 40, 100$ measurements and $d_Y = 0$. %(b) MSE of estimator $\check d_Y^\mathrm{(SLI)}$ as a function of the separation in $y$-direction for $L = 20, 40, 100$ measurements and $d_X = 0$.
}
\end{figure}

\begin{figure}
\includegraphics[width=3.5in]{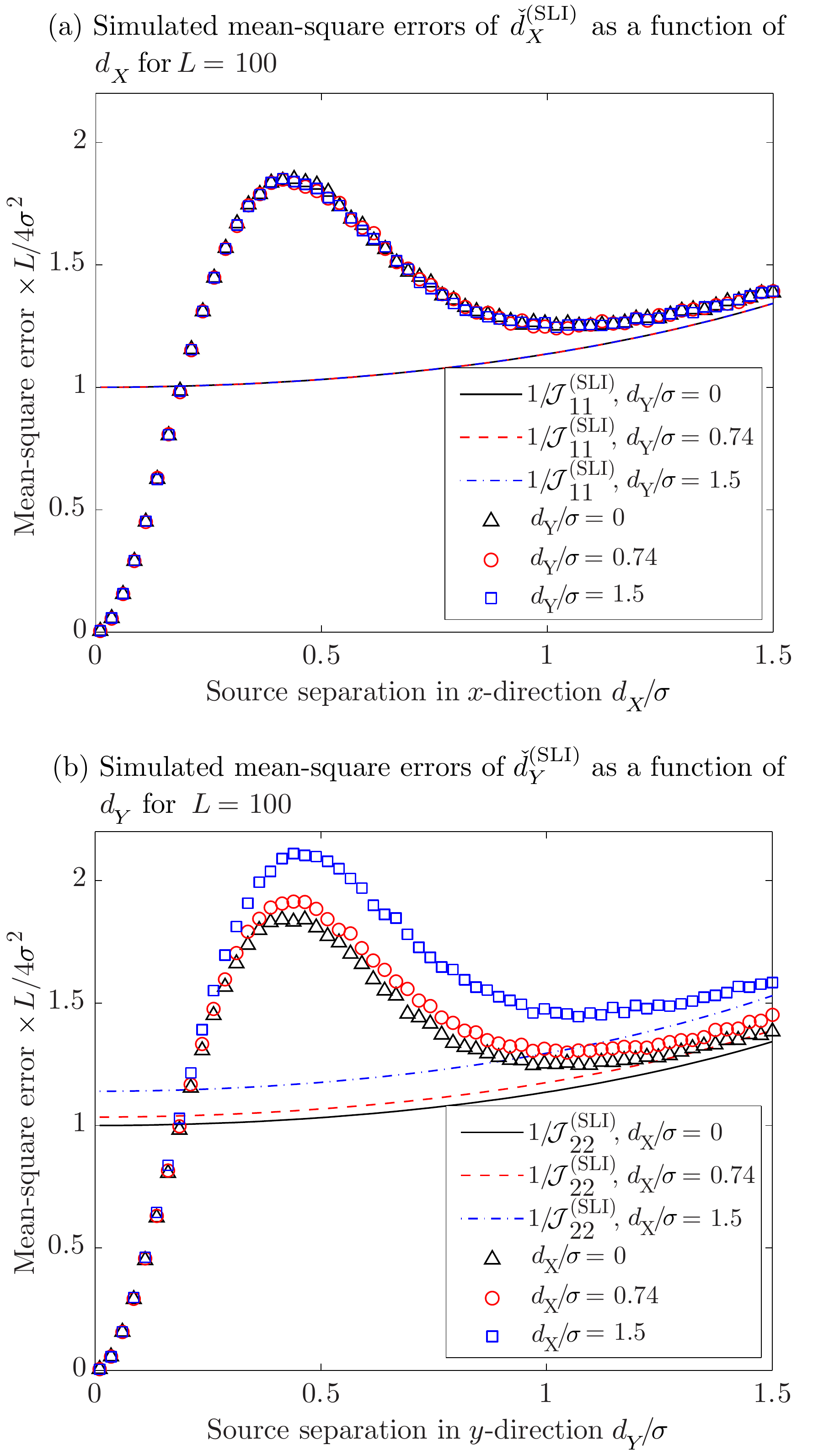}
\caption{\label{fig:MSE_SLI_var}Simulated mean-square errors of SLIVER with maximum likelihood estimators of Eq.~\eqref{eq:ML_SLI} for different values of $d_X$ and $d_Y$ in $L=100$ trials. The corresponding Cram\'er-Rao bounds are included in the plots for comparison. (a)~MSE of estimator $\check d_X^\mathrm{(SLI)}$ as a function of the separation in $x$-direction for $d_Y/\sigma = 0, 0.74, 1.5$. (b)~MSE of estimator $\check d_Y^\mathrm{(SLI)}$ as a function of the separation in $y$-direction for $d_X/\sigma = 0, 0.74, 1.5$.}
\end{figure}

Figures~\ref{fig:MSE_SLI} and \ref{fig:MSE_SLI_var} show the simulated MSEs of the ML estimators in Eq.~\eqref{eq:ML_SLI}. The plotted MSEs are scaled relative to the  value of the QCR bound for that $L$. 
Fig.~\ref{fig:MSE_SLI} plots the MSE of  $\check d_X^\mathrm{(SLI)}$ as a function of $x$-separation for $d_Y = 0$.The MSE of estimator $\check d_Y^\mathrm{(SLI)}$ as a function of $y$-separation for $d_X = 0$ is virtually identical to that of Fig.~\ref{fig:MSE_SLI} and is not shown. %This behavior is expected as $E_S(x, y) = E(x,y)$ if $d_X=0$.
We see that the ML estimator beats the CR bound for small $d_X/\sigma$ due to the biasedness of the estimator placing it beyond the purview of the CR bound. This ``super-efficiency'' effect is well-known in classical estimation. We refer the reader to Appendix E of ref.~\cite{tsang16a} and in particular ref.~\cite{Tsa16} for extensive discussion on this point. Here we just note that the range of $d_X$ values where the estimation is super-efficient  shrinks with increasing $L$ and limits its practical usefulness.  

Fig.~\ref{fig:MSE_SLI_var} explores the effect of nonzero values of the separations $d_Y (d_X)$ on the MSE of  $\check d_X^\mathrm{(SLI)} ( \check d_Y^\mathrm{(SLI)})$ -- the number of measurements is fixed at $L = 100$  and the corresponding CR bounds for the relevant separations are shown. Fig.~\ref{fig:MSE_SLI_var}(a) shows the simulated MSE of estimator $\check d_X^\mathrm{(SLI)}$ as a function of $d_X$, for $d_Y = 0, 0.74\sigma,$ and  $1.5\sigma$. We see that both the estimator and the CR bound show  little dependence on the separation $d_Y$. Fig.~\ref{fig:MSE_SLI_var}(b) plots the simulated MSE of estimator $\check d_Y^\mathrm{(SLI)}$ against $d_Y$ for the case of $d_X = 0, 0.74\sigma,$ and  $1.5\sigma$. As $d_X$ increases, the CR bound increases along with the MSE of the estimator.

\subsection{Monte-Carlo analysis of SPADE with maximum-likelihood estimation}

\begin{figure}
\includegraphics[width=2.5in]{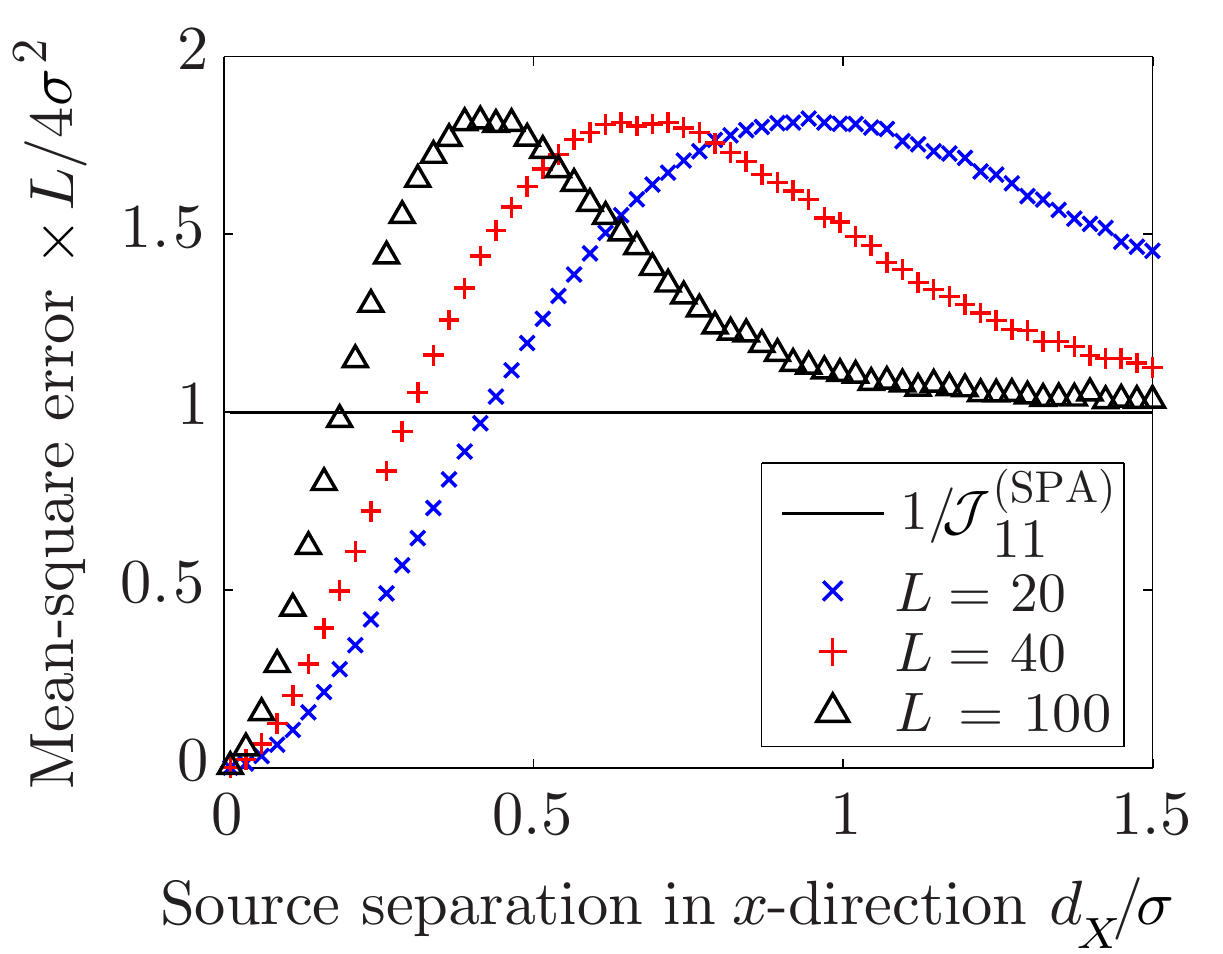}\\
\caption{\label{fig:MSE_SPA}Simulated mean-square errors of SPADE with maximum likelihood estimators of Eq.~\eqref{eq:ML_SPA}. (a) MSE of estimator $\check d_X^\mathrm{(SPA)}$ as a function of the separation in $x$-direction for $L = 20, 40, 100$ measurements. The MSE behavior of $\check d_Y^\mathrm{(SPA)}$ as a function of the $y$-separation is similar. }
\end{figure}

Given $L$ photon detections in the SPADE system of Sec.~\ref{sec:spade}, the post-selected measurement record consists of a sequence $\{(q_l,r_l)\}_{l=1}^L$ of indices of the 2-D HG modes in which the $L$ photons were detected.  The ML estimators for $d_X$ and $d_Y$ can be shown to be
\begin{align}\label{eq:ML_SPA}
\check d_{X}^\mathrm{(SPA)} = 4\sigma\sqrt{\frac{H_X}{L}}, \qquad
\check d_{Y}^\mathrm{(SPA)} = 4\sigma\sqrt{\frac{H_Y}{L}}.
\end{align}
where $H_X = \sum_{l=1}^L q_l$ and $H_Y = \sum_{l=1}^L r_l$. Fig.~\ref{fig:MSE_SPA} displays the results for the MSE of $\check d_{X}^\mathrm{(SPA)}$. The MSE behavior of $\check d_{Y}^\mathrm{(SPA)}$ is identical and is not shown. The performance of $\check d_{X}^\mathrm{(SPA)} (\check d_{Y}^\mathrm{(SPA)})$ is independent of the value of $d_Y (d_X)$. These behaviors are expected from Eq.~\eqref{eq:spadeprob} -- The joint probability of $q$ and $r$ is a product of their marginal distributions that respectively depend only on $d_X$ and $d_Y$ in identical fashion. The ML estimators beat the CR bounds in the estimation of $d_X$ and $d_Y$ for small separations. The errors remain less than twice the CR bounds for any separations.

\section{Discussion and Outlook}\label{sec:conclusion}
In this paper, we have calculated the QCR bound for locating two weak incoherent optical point sources on a two-dimensional plane using an imaging system with any given inversion-symmetric point-spread function. The key result is that, in stark contrast to spatially-resolved direct imaging \cite{ram06,tsang16a}, the bounds on the MSE of estimating the $x$- and $y$-separations are independent  of the vector separation between the two sources. Strictly speaking, a large enough separation can take us outside the scalar-field paraxial approximation assumed here, but that approximation is excellent for most practical applications in microscopy and telescopy.

We have also proposed and analyzed two measurement schemes -- the extended SLIVER and SPADE schemes -- for simultaneously estimating the components of the separation, whose classical CR bounds approach the quantum bounds for sub-Rayleigh separations (SLIVER) or all separations if the PSF is Gaussian (SPADE). Monte-Carlo simulations show that the two schemes have MSEs no larger than twice predicted by the quantum limits for values of  source separation from zero to beyond the PSF width.

The extended SLIVER scheme given here does not employ an image inversion device (see, e.g., refs.~\cite{WSH09,WBK+11} in which the general properties of the device were studied and ref.~\cite{SDL16} for an implementation in the context of two-source resolution), which was required in the scheme of ref.~\cite{nair16}  tailored to  directly estimate  the magnitude $d = \sqrt{d_X^2 + d_Y^2}$ of the separation. Thus, each interferometer stage in the current scheme may be technically simpler to implement than that of ref.~\cite{nair16}, and especially if only one component of the separation is of interest. However, the original scheme is likely to be superior for estimation of $d = \sqrt{d_X^2 + d_Y^2}$, as suggested by the dependence of the MSE of  $\check d_Y^\mathrm{(SLI)}$ on $d_X$ in the simulations in Sec.~\ref{sec:analysis}.

Our analysis here can be extended in various directions. By adopting the Gaussian-state source model of ref.~\cite{NT16Gaussian},  our quantum Fisher information calculations can in principle be extended to sources of arbitrary strength. The study of two-source transverse localization can also be generalized to sources emitting light in more general quantum states, as in ref.~\cite{lupo16}. In principle, it can also be extended to multiple sources, although finding near-optimal measurement schemes is likely to be challenging. The performance of the SPADE and SLIVER schemes can also be analyzed in the thermal-state model in the spirit of ref.~\cite{nair16}.
Even with the same source model as here, the performance of SPADE employing only a finite number of HG modes and extensions of the SLIVER scheme using pixelated detectors (cf. the fin-SPADE and pix-SLIVER  schemes of \cite{NT16Gaussian}) can be explored, as well as generalizations enabling full transverse localization of the sources. This problem will be explored in the future.   

 In both measurement schemes, we have assumed that the centroid position is known. If that knowledge is unavailable, a portion of the light can be used for image-plane photon counting to determine the centroid position before performing either of the schemes as detailed in ref.~\cite{tsang16a}. To account for the residual error in estimating the centroid, it is important to study the performance of SLIVER and SPADE if the centroid is not aligned to the optical axis. Among the other practical questions to be explored are the effects of non-unity coupling efficiency in SPADE, and  unequal detection efficiencies of the detectors.

\begin{center}
\line(1,0){200}
\end{center}
%%%%%%%%%%%%%

\appendix
\section{Quantum Fisher Information Matrix} \label{sec:QFI}
To evaluate the QFI matrix, we first need to diagonalize $\rho$ including enough eigenvectors to span the combined support of $\rho$ and the $\left\{\partial\rho/\partial \theta_\mu\right\}$. The partial derivatives of $\rho$ with respect to the object-plane source coordinates $X_\mu$ and $Y_\mu$ are
\allowdisplaybreaks{
\begin{align}
\frac{\partial\rho}{\partial X_\mu} &= \frac{\partial D_1}{\partial X_\mu}|e_1\rangle\langle e_1|+\frac{\partial D_2}{\partial X_\mu}|e_2\rangle\langle e_2| \nonumber\\
&\quad + \left( D_1\frac{\partial |e_1\rangle}{\partial X_\mu}\langle e_1|+D_2\frac{\partial |e_2\rangle}{\partial X_\mu}\langle e_2| + \mathrm{H.c} \right), \label{eq:derivativeX} \\
\frac{\partial\rho}{\partial Y_\mu} &= \frac{\partial D_1}{\partial Y_\mu}|e_1\rangle\langle e_1|+\frac{\partial D_2}{\partial Y_\mu}|e_2\rangle\langle e_2| \nonumber\\
&\quad + \left( D_1\frac{\partial |e_1\rangle}{\partial Y_\mu}\langle e_1|+D_2\frac{\partial |e_2\rangle}{\partial Y_\mu}\langle e_2| + \mathrm{H.c.} \right), \label{eq:derivativeY} 
\end{align}
}where H.c. denotes the Hermitian conjugate.
For any (possibly complex-valued) $\psi(x,y)$ symmetric about the origin, viz.,
\begin{equation}
\psi(x,y) = \psi(-x,-y),
\end{equation}
here we show that the overlap $\delta$  given by Eq. \eqref{eq:delta} is real-valued. The complex conjugate of $\delta$ is given by
\begin{align}
\delta^* &= \int_{-\infty}^\infty \mathrm{d}x \int_{-\infty}^\infty \mathrm{d}y \,\psi(x-X_1, y-Y_1)\psi^*(x-X_2, y-Y_2).
\end{align}
Apply the transformations
\begin{align}
x &\mapsto \frac{\bar X}{2} -x, \qquad y \mapsto \frac{\bar Y}{2} - y, 
\end{align}
and flip the limits of both integrations, we have
\begin{align}
\delta^* &= \int_{-\infty}^\infty \mathrm{d}x \int_{-\infty}^\infty \mathrm{d}y \,\psi\left(-x+\frac{d_X}{2}, -y+\frac{d_Y}{2}\right) \nonumber\\
&\qquad \times \psi^*\left(-x-\frac{d_X}{2},-y-\frac{d_Y}{2}\right),
\end{align}
where $\bar X, \bar Y, d_X$ and $d_Y$ are defined in Eqs.~\eqref{eq:barXbarY} and \eqref{eq:dXdY}. Using the symmetricity of $\psi(x,y)$,
\begin{align}
\delta^* &= \int_{-\infty}^\infty \mathrm{d}x \int_{-\infty}^\infty \mathrm{d}y \,\psi\left(x-\frac{d_X}{2}, y-\frac{d_Y}{2}\right) \nonumber\\
&\qquad \times \psi^*\left(x+\frac{d_X}{2},y+\frac{d_Y}{2}\right) \nonumber\\
&=\int_{-\infty}^\infty \mathrm{d}x \int_{-\infty}^\infty \mathrm{d}y \,\psi(x-X_2, y-Y_2)\psi^*(x-X_1, y-Y_1) \nonumber\\
&= \delta,
\end{align}
where we apply the transformations
\begin{align}
x &\mapsto x -\frac{\bar X}{2}, \qquad y \mapsto y - \frac{\bar Y}{2}, 
\end{align}
in the second equality. Hence, $\delta$ is real-valued.

After some algebra it can be shown that a possible set of eigenvectors of $\rho$  is 
\begin{widetext}
{\allowdisplaybreaks
\begin{align} \label{eq:basis}
|e_0\rangle &= |\mathrm{vac}\rangle, \qquad 
|e_1\rangle = \frac{1}{\sqrt{2(1-\delta)}}(|\psi_1\rangle - |\psi_2\rangle), \qquad
|e_2\rangle = \frac{1}{\sqrt{2(1+\delta)}}(|\psi_1\rangle + |\psi_2\rangle), \nonumber\\
|e_3\rangle &= \frac{1}{c_3} \left[\Delta k_X(|\psi_{1X}\rangle+|\psi_{2X}\rangle)+r_+\Delta k_Y(|\psi_{1Y}\rangle+|\psi_{2Y}\rangle) -\frac{2(\gamma_X+r_+\gamma_Y)}{\sqrt{2(1-\delta)}}|e_1\rangle\right], \nonumber\\
|e_4\rangle &= \frac{1}{c_4} \left[\Delta k_X(|\psi_{1X}\rangle+|\psi_{2X}\rangle)-r_+\Delta k_Y(|\psi_{1Y}\rangle+|\psi_{2Y}\rangle) -\frac{2(\gamma_X-r_+\gamma_Y)}{\sqrt{2(1-\delta)}}|e_1\rangle\right], \nonumber\\
|e_5\rangle &= \frac{1}{c_5} \left[\Delta k_X(|\psi_{1X}\rangle-|\psi_{2X}\rangle)+r_-\Delta k_Y(|\psi_{1Y}\rangle-|\psi_{2Y}\rangle) +\frac{2(\gamma_X+r_-\gamma_Y)}{\sqrt{2(1+\delta)}}|e_2\rangle\right], \nonumber\\
|e_6\rangle &= \frac{1}{c_6} \left[\Delta k_X(|\psi_{1X}\rangle-|\psi_{2X}\rangle)-r_-\Delta k_Y(|\psi_{1Y}\rangle-|\psi_{2Y}\rangle) +\frac{2(\gamma_X-r_-\gamma_Y)}{\sqrt{2(1+\delta)}}|e_2\rangle\right],
\end{align}
}
\end{widetext}
where $\delta$ is given by Eq. \eqref{eq:delta}, $\Delta k_X, \Delta k_Y, \gamma_X, \gamma_Y$ are defined in Eq. \eqref{eq:parameter},
{\allowdisplaybreaks
\begin{align}
|\psi_{1X}\rangle &\equiv \frac{1}{\Delta k_X}\int_{-\infty}^\infty \mathrm dx \int_{-\infty}^\infty \mathrm dy \frac{\partial\psi(x-X_1, y-Y_1)}{\partial X_1}|x,y \rangle, \nonumber\\
|\psi_{2X}\rangle &\equiv \frac{1}{\Delta k_X}\int_{-\infty}^\infty \mathrm dx \int_{-\infty}^\infty \mathrm dy \frac{\partial\psi(x-X_2, y-Y_2)}{\partial X_2}|x,y \rangle, \nonumber\\
|\psi_{1Y}\rangle &\equiv \frac{1}{\Delta k_Y}\int_{-\infty}^\infty \mathrm dx \int_{-\infty}^\infty \mathrm dy \frac{\partial\psi(x-X_1, y-Y_1)}{\partial Y_1}|x,y \rangle, \nonumber\\
|\psi_{2Y}\rangle &\equiv \frac{1}{\Delta k_Y}\int_{-\infty}^\infty \mathrm dx \int_{-\infty}^\infty \mathrm dy \frac{\partial\psi(x-X_2, y-Y_2)}{\partial Y_2}|x,y \rangle, \nonumber\\
c_3 &\equiv 2\sqrt{\Delta k_X^2+b_X^2 -\frac{\gamma_X^2}{1-\delta}+|r_+|\left|a+a_s-\frac{\gamma_X\gamma_Y}{1-\delta}\right|}, \nonumber\\
c_4 &\equiv 2\sqrt{\Delta k_X^2+b_X^2 -\frac{\gamma_X^2}{1-\delta}-|r_+|\left|a+a_s-\frac{\gamma_X\gamma_Y}{1-\delta}\right|}, \nonumber\\
c_5 &\equiv 2\sqrt{\Delta k_X^2-b_X^2 -\frac{\gamma_X^2}{1+\delta}+|r_-|\left|a-a_s-\frac{\gamma_X\gamma_Y}{1+\delta}\right|}, \nonumber\\
c_6 &\equiv 2\sqrt{\Delta k_X^2-b_X^2 -\frac{\gamma_X^2}{1+\delta}-|r_-|\left|a-a_s-\frac{\gamma_X\gamma_Y}{1+\delta}\right|}, \nonumber\\
r_\pm &\equiv \left[\frac{\Delta k_X^2 \pm b_X^2-\gamma_X^2/(1\mp\delta)}{\Delta k_Y^2+b_Y^2-\gamma_Y^2/(1\mp\delta)}\right]^{1/2}\nonumber\\
&\quad \times \exp\left[-i\arg \left(a \pm a_s-\frac{\gamma_X\gamma_Y}{1\mp\delta}\right)\right], \nonumber\\
a &\equiv \int_{-\infty}^\infty \mathrm dx \int_{-\infty}^\infty \mathrm dy \frac{\partial \psi^*(x,y)}{\partial x} \frac{\partial \psi(x,y)}{\partial y}, \nonumber\\
a_s &\equiv \int_{-\infty}^\infty \mathrm dx \int_{-\infty}^\infty \mathrm dy \frac{\partial \psi^*(x,y)}{\partial x} \frac{\partial \psi(x-d_X,y-d_Y)}{\partial y}, \nonumber\\
b_X^2 &\equiv \int_{-\infty}^\infty \mathrm dx \int_{-\infty}^\infty \mathrm dy \left[\frac{\partial \psi^*(x-X_1,y-Y_1)}{\partial X_1}\right. \nonumber\\
&\quad \left. \times \frac{\partial \psi(x-X_2,y-Y_2)}{\partial X_2} \right], \nonumber\\
b_Y^2 &\equiv \int_{-\infty}^\infty \mathrm dx \int_{-\infty}^\infty \mathrm dy \left[\frac{\partial \psi^*(x-X_1,y-Y_1)}{\partial Y_1}\right. \nonumber\\
&\quad \left. \times \frac{\partial \psi(x-X_2,y-Y_2)}{\partial Y_2} \right],
\end{align}
}
and the eigenvalues of $\rho$ ($D_n$ corresponding to $|e_n\rangle$) are
\begin{align}
D_0 &= 1-\epsilon, \quad D_1 = \frac{\epsilon}{2}(1-\delta), \quad D_2 = \frac{\epsilon}{2}(1+\delta), \nonumber\\
D_3 &= D_4 = D_5 = D_6 =0.
\end{align}

The SLDs with respect to the derivative in Eqs. \eqref{eq:derivativeX} and \eqref{eq:derivativeY} can be found using Eq. \eqref{eq:SLD},
\begin{align}
\mathcal L_\mu^{(X)} &= \sum_{\substack{m,n\\D_m+D_n\neq 0}} \frac{2}{D_m+D_n}\langle e_m|\frac{\partial\rho}{\partial X_\mu}|e_n\rangle |e_m\rangle\langle e_n|, \nonumber\\
\mathcal L_\mu^{(Y)} &= \sum_{\substack{m,n\\D_m+D_n\neq 0}} \frac{2}{D_m+D_n}\langle e_m|\frac{\partial\rho}{\partial Y_\mu}|e_n\rangle |e_m\rangle\langle e_n|.
\end{align}
Transforming to  the centroid and separation parameters $\theta$ of Eq. \eqref{eq:theta} gives the SLDs 
\begin{align}
\mathcal L_1 &= \mathcal L_1^{(X)}+\mathcal L_2^{(X)}, \qquad \mathcal L_2 = \mathcal L_1^{(Y)}+\mathcal L_2^{(Y)}, \nonumber\\
\mathcal L_3 &= \frac{\mathcal L_2^{(X)}-\mathcal L_1^{(X)}}{2}, \qquad \mathcal L_4 = \frac{\mathcal L_2^{(Y)}-\mathcal L_1^{(Y)}}{2}.
\end{align}
We can now evaluate the quantum Fisher information using Eq.~\eqref{eq:qfi} to finally obtain Eq.~\eqref{eq:J}.

\section{The statistics of SLIVER} \label{sec:SLIVER}

In this Appendix, we compute the probabilities (Eq.~\eqref{eq:Pr})
\begin{align} \label{eq:appPr}
P(r) &= \operatorname{tr}(\rho \hat{N}_r); \;\; r=1,2,3, \\
&= \sum_{s=1,2} \epsilon_s \langle \Psi_s| \hat{N}_r| \Psi_s \rangle
\end{align} 
of the three possible cases of detecting one photon in the SLIVER measurement. Here,  the states
\begin{align}
\ket{\Psi_s} = \ket{\psi_s}\ket{0}_1\ket{0}_2\;;s=1,2,
\end{align}
as the single-photon source states augmented with vacuum states in the extra beam-splitter input modes $\hat{V}_1(x,y)$ and $\hat{V}_2(x,y)$.
From Eq.~\eqref{eq:Nrdef} and Eqs.~\eqref{eq:E_X}-\eqref{eq:E_R},  knowledge of the second moments
\begin{align}
&\langle\Psi_s|\hat{E}^\dagger(x,y)\hat{E}(x',y')|\Psi_s\rangle, \label{Kn}\\
&\langle\Psi_s|\hat{E}^\dagger(x,y)\hat{V}_1(x',y')|\Psi_s\rangle, \\
&\langle\Psi_s|\hat{E}^\dagger(x,y)\hat{V}_2(x',y')|\Psi_s\rangle, \\
&\langle\Psi_s|\hat{V}_1^\dagger(x,y)\hat{V}_1(x',y')|\Psi_s\rangle, \\
&\langle\Psi_s|\hat{V}_1^\dagger(x,y)\hat{V}_2(x',y')|\Psi_s\rangle, \\
&\langle\Psi_s|\hat{V}_2^\dagger(x,y)\hat{V}_2(x',y')|\Psi_s\rangle 
\end{align}
for arbitrary $(x,y)$ and $(x',y')$ and for $s=1,2$, suffices to calculate Eq.~\eqref{eq:appPr}. Since $\hat{V}_1(x,y)$ and $\hat{V}_2(x,y)$ are in vacuum, the only nonzero second moment is Eq.~\eqref{Kn}. To calculate $\langle\Psi_s|\hat{E}^\dagger(x,y)\hat{E}(x',y')|\Psi_s\rangle$, expand the image plane field in terms of a complete orthonormal set $\{ \varphi_q(x,y)\}_{q=0}^\infty$ with associated annihilation operators $\{\hat{a}_q\}_{q=0}^\infty$ such that
$\varphi_0(x,y) = \psi_1(x,y)$:
\begin{align} \label{eq:E1}
\hat{E}(x,y) &= \hat{a}_0\psi_1(x,y)+\sum_{q=1}^\infty \hat{a}_q\varphi_q(x,y).
\end{align}
Then, by definition of the single-photon state \eqref{eq:state_psi}, the mode $\hat{a}_0$ is in a single photon state while 
 $\{\hat{a}_q\}_{q=1}^\infty$ are all in vacuum. It follows that 
\begin{align}
&\langle\Psi_1|\hat{E}^\dagger(x,y)\hat{E}(x',y')|\Psi_1\rangle \\
&= \bra{1}{\hat{a}^\dag_0 \hat{a}_0}\ket{1}\,\psi_1^*(x,y)\psi_1(x',y') \\
&= \psi_1^*(x,y)\psi_1(x',y').
\end{align} Using a similar mode expansion for $s=2$, we have generally
\begin{align}
\langle\Psi_s|\hat{E}^\dagger(x,y)\hat{E}(x',y')|\Psi_s \rangle = \psi_s^*(x,y)\,\psi_s(x',y') \;\; ;s=1,2.
\end{align}
Using this along with Eqs.~\eqref{eq:E_X}-\eqref{eq:E_R} to evaluate Eq.~\eqref{eq:appPr} results in Eqs.~\eqref{eq:Pr_sol} of Sec.~\ref{sec:sliver}.

\begin{acknowledgments}
S.~Z.~Ang, R. Nair and M.~Tsang acknowledge support by the Singapore National Research Foundation under NRF Grant.~No.~NRF-NRFF2011-07 and the Singapore Ministry of Education Academic Research Fund Tier 1 Project R-263-000-C06-112.
\end{acknowledgments}


\begin{thebibliography}{40}%
\makeatletter
\providecommand \@ifxundefined [1]{%
 \@ifx{#1\undefined}
}%
\providecommand \@ifnum [1]{%
 \ifnum #1\expandafter \@firstoftwo
 \else \expandafter \@secondoftwo
 \fi
}%
\providecommand \@ifx [1]{%
 \ifx #1\expandafter \@firstoftwo
 \else \expandafter \@secondoftwo
 \fi
}%
\providecommand \natexlab [1]{#1}%
\providecommand \enquote  [1]{``#1''}%
\providecommand \bibnamefont  [1]{#1}%
\providecommand \bibfnamefont [1]{#1}%
\providecommand \citenamefont [1]{#1}%
\providecommand \href@noop [0]{\@secondoftwo}%
\providecommand \href [0]{\begingroup \@sanitize@url \@href}%
\providecommand \@href[1]{\@@startlink{#1}\@@href}%
\providecommand \@@href[1]{\endgroup#1\@@endlink}%
\providecommand \@sanitize@url [0]{\catcode `\\12\catcode `\$12\catcode
  `\&12\catcode `\#12\catcode `\^12\catcode `\_12\catcode `\%12\relax}%
\providecommand \@@startlink[1]{}%
\providecommand \@@endlink[0]{}%
\providecommand \url  [0]{\begingroup\@sanitize@url \@url }%
\providecommand \@url [1]{\endgroup\@href {#1}{\urlprefix }}%
\providecommand \urlprefix  [0]{URL }%
\providecommand \Eprint [0]{\href }%
\providecommand \doibase [0]{http://dx.doi.org/}%
\providecommand \selectlanguage [0]{\@gobble}%
\providecommand \bibinfo  [0]{\@secondoftwo}%
\providecommand \bibfield  [0]{\@secondoftwo}%
\providecommand \translation [1]{[#1]}%
\providecommand \BibitemOpen [0]{}%
\providecommand \bibitemStop [0]{}%
\providecommand \bibitemNoStop [0]{.\EOS\space}%
\providecommand \EOS [0]{\spacefactor3000\relax}%
\providecommand \BibitemShut  [1]{\csname bibitem#1\endcsname}%
\let\auto@bib@innerbib\@empty
%</preamble>
\bibitem [{\citenamefont {{Lord Rayleigh}}(1879)}]{rayleigh1879xxxi}%
  \BibitemOpen
  \bibfield  {author} {\bibinfo {author} {\bibnamefont {{Lord Rayleigh}}},\
  }\href {\doibase 10.1080/14786447908639684} {\bibfield  {journal} {\bibinfo
  {journal} {The London, Edinburgh, and Dublin Philosophical Magazine and
  Journal of Science}\ }\textbf {\bibinfo {volume} {8}},\ \bibinfo {pages}
  {261} (\bibinfo {year} {1879})}\BibitemShut {NoStop}%
\bibitem [{\citenamefont {Bettens}\ \emph {et~al.}(1999)\citenamefont
  {Bettens}, \citenamefont {Dyck}, \citenamefont {den Dekker}, \citenamefont
  {Sijbers},\ and\ \citenamefont {van~den Bos}}]{bettens99}%
  \BibitemOpen
  \bibfield  {author} {\bibinfo {author} {\bibfnamefont {E.}~\bibnamefont
  {Bettens}}, \bibinfo {author} {\bibfnamefont {D.~V.}\ \bibnamefont {Dyck}},
  \bibinfo {author} {\bibfnamefont {A.}~\bibnamefont {den Dekker}}, \bibinfo
  {author} {\bibfnamefont {J.}~\bibnamefont {Sijbers}}, \ and\ \bibinfo
  {author} {\bibfnamefont {A.}~\bibnamefont {van~den Bos}},\ }\href {\doibase
  http://dx.doi.org/10.1016/S0304-3991(99)00006-6} {\bibfield  {journal}
  {\bibinfo  {journal} {Ultramicroscopy}\ }\textbf {\bibinfo {volume} {77}},\
  \bibinfo {pages} {37 } (\bibinfo {year} {1999})}\BibitemShut {NoStop}%
\bibitem [{\citenamefont {Aert}\ \emph {et~al.}(2002)\citenamefont {Aert},
  \citenamefont {den Dekker}, \citenamefont {Dyck},\ and\ \citenamefont
  {van~den Bos}}]{vanaert02}%
  \BibitemOpen
  \bibfield  {author} {\bibinfo {author} {\bibfnamefont {S.~V.}\ \bibnamefont
  {Aert}}, \bibinfo {author} {\bibfnamefont {A.}~\bibnamefont {den Dekker}},
  \bibinfo {author} {\bibfnamefont {D.~V.}\ \bibnamefont {Dyck}}, \ and\
  \bibinfo {author} {\bibfnamefont {A.}~\bibnamefont {van~den Bos}},\ }\href
  {\doibase http://dx.doi.org/10.1016/S1047-8477(02)00016-3} {\bibfield
  {journal} {\bibinfo  {journal} {Journal of Structural Biology}\ }\textbf
  {\bibinfo {volume} {138}},\ \bibinfo {pages} {21 } (\bibinfo {year}
  {2002})}\BibitemShut {NoStop}%
\bibitem [{\citenamefont {Ram}\ \emph {et~al.}(2006)\citenamefont {Ram},
  \citenamefont {Ward},\ and\ \citenamefont {Ober}}]{ram06}%
  \BibitemOpen
  \bibfield  {author} {\bibinfo {author} {\bibfnamefont {S.}~\bibnamefont
  {Ram}}, \bibinfo {author} {\bibfnamefont {E.~S.}\ \bibnamefont {Ward}}, \
  and\ \bibinfo {author} {\bibfnamefont {R.~J.}\ \bibnamefont {Ober}},\ }\href
  {\doibase 10.1073/pnas.0508047103} {\bibfield  {journal} {\bibinfo  {journal}
  {Proceedings of the National Academy of Sciences of the United States of
  America}\ }\textbf {\bibinfo {volume} {103}},\ \bibinfo {pages} {4457}
  (\bibinfo {year} {2006})}\BibitemShut {NoStop}%
\bibitem [{\citenamefont {Van~Trees}(2001)}]{vantrees01}%
  \BibitemOpen
  \bibfield  {author} {\bibinfo {author} {\bibfnamefont {H.~L.}\ \bibnamefont
  {Van~Trees}},\ }\href@noop {} {\emph {\bibinfo {title} {Detection,
  Estimation, and Modulation Theory Part. I}}}\ (\bibinfo  {publisher} {John
  Wiley \& Sons, New York},\ \bibinfo {year} {2001})\BibitemShut {NoStop}%
\bibitem [{\citenamefont {Betzig}(1995)}]{betzig95}%
  \BibitemOpen
  \bibfield  {author} {\bibinfo {author} {\bibfnamefont {E.}~\bibnamefont
  {Betzig}},\ }\href {\doibase 10.1364/OL.20.000237} {\bibfield  {journal}
  {\bibinfo  {journal} {Opt. Lett.}\ }\textbf {\bibinfo {volume} {20}},\
  \bibinfo {pages} {237} (\bibinfo {year} {1995})}\BibitemShut {NoStop}%
\bibitem [{\citenamefont {Moerner}\ and\ \citenamefont
  {Kador}(1989)}]{moerner89}%
  \BibitemOpen
  \bibfield  {author} {\bibinfo {author} {\bibfnamefont {W.~E.}\ \bibnamefont
  {Moerner}}\ and\ \bibinfo {author} {\bibfnamefont {L.}~\bibnamefont
  {Kador}},\ }\href {\doibase 10.1103/PhysRevLett.62.2535} {\bibfield
  {journal} {\bibinfo  {journal} {Phys. Rev. Lett.}\ }\textbf {\bibinfo
  {volume} {62}},\ \bibinfo {pages} {2535} (\bibinfo {year}
  {1989})}\BibitemShut {NoStop}%
\bibitem [{\citenamefont {Hell}\ and\ \citenamefont {Wichmann}(1994)}]{hell94}%
  \BibitemOpen
  \bibfield  {author} {\bibinfo {author} {\bibfnamefont {S.~W.}\ \bibnamefont
  {Hell}}\ and\ \bibinfo {author} {\bibfnamefont {J.}~\bibnamefont
  {Wichmann}},\ }\href {\doibase 10.1364/OL.19.000780} {\bibfield  {journal}
  {\bibinfo  {journal} {Opt. Lett.}\ }\textbf {\bibinfo {volume} {19}},\
  \bibinfo {pages} {780} (\bibinfo {year} {1994})}\BibitemShut {NoStop}%
\bibitem [{\citenamefont {Weisenburger}\ and\ \citenamefont
  {Sandoghdar}(2015)}]{WS15}%
  \BibitemOpen
  \bibfield  {author} {\bibinfo {author} {\bibfnamefont {S.}~\bibnamefont
  {Weisenburger}}\ and\ \bibinfo {author} {\bibfnamefont {V.}~\bibnamefont
  {Sandoghdar}},\ }\href {\doibase 10.1080/00107514.2015.1026557} {\bibfield
  {journal} {\bibinfo  {journal} {Contemporary Physics}\ }\textbf {\bibinfo
  {volume} {56}},\ \bibinfo {pages} {123} (\bibinfo {year} {2015})}\BibitemShut
  {NoStop}%
\bibitem [{\citenamefont {Helstrom}(1976)}]{helstrom76}%
  \BibitemOpen
  \bibfield  {author} {\bibinfo {author} {\bibfnamefont {C.~W.}\ \bibnamefont
  {Helstrom}},\ }\href@noop {} {\emph {\bibinfo {title} {Quantum Detection and
  Estimation Theory}}}\ (\bibinfo  {publisher} {Academic Press, New York},\
  \bibinfo {year} {1976})\BibitemShut {NoStop}%
\bibitem [{\citenamefont {Holevo}(2011)}]{holevo11}%
  \BibitemOpen
  \bibfield  {author} {\bibinfo {author} {\bibfnamefont {A.~S.}\ \bibnamefont
  {Holevo}},\ }\href@noop {} {\emph {\bibinfo {title} {Statistical Structure of
  Quantum Theory}}}\ (\bibinfo  {publisher} {Springer Berlin Heidelberg},\
  \bibinfo {year} {2011})\BibitemShut {NoStop}%
\bibitem [{\citenamefont {Paris}(2009)}]{paris09}%
  \BibitemOpen
  \bibfield  {author} {\bibinfo {author} {\bibfnamefont {M.~G.~A.}\
  \bibnamefont {Paris}},\ }\href {\doibase 10.1142/S0219749909004839}
  {\bibfield  {journal} {\bibinfo  {journal} {International Journal of Quantum
  Information}\ }\textbf {\bibinfo {volume} {07}},\ \bibinfo {pages} {125}
  (\bibinfo {year} {2009})}\BibitemShut {NoStop}%
\bibitem [{\citenamefont {Tsang}(2015)}]{tsang15b}%
  \BibitemOpen
  \bibfield  {author} {\bibinfo {author} {\bibfnamefont {M.}~\bibnamefont
  {Tsang}},\ }\href {\doibase 10.1364/OPTICA.2.000646} {\bibfield  {journal}
  {\bibinfo  {journal} {Optica}\ }\textbf {\bibinfo {volume} {2}},\ \bibinfo
  {pages} {646} (\bibinfo {year} {2015})}\BibitemShut {NoStop}%
\bibitem [{\citenamefont {Tsang}\ \emph
  {et~al.}(2016{\natexlab{a}})\citenamefont {Tsang}, \citenamefont {Nair},\
  and\ \citenamefont {Lu}}]{tsang16a}%
  \BibitemOpen
  \bibfield  {author} {\bibinfo {author} {\bibfnamefont {M.}~\bibnamefont
  {Tsang}}, \bibinfo {author} {\bibfnamefont {R.}~\bibnamefont {Nair}}, \ and\
  \bibinfo {author} {\bibfnamefont {X.-M.}\ \bibnamefont {Lu}},\ }\href
  {\doibase 10.1103/PhysRevX.6.031033} {\bibfield  {journal} {\bibinfo
  {journal} {Physical Review X}\ }\textbf {\bibinfo {volume} {6}},\ \bibinfo
  {pages} {031033} (\bibinfo {year} {2016}{\natexlab{a}})}\BibitemShut
  {NoStop}%
\bibitem [{\citenamefont {Tsang}\ \emph
  {et~al.}(2016{\natexlab{b}})\citenamefont {Tsang}, \citenamefont {Nair},\
  and\ \citenamefont {Lu}}]{tsang16b}%
  \BibitemOpen
  \bibfield  {author} {\bibinfo {author} {\bibfnamefont {M.}~\bibnamefont
  {Tsang}}, \bibinfo {author} {\bibfnamefont {R.}~\bibnamefont {Nair}}, \ and\
  \bibinfo {author} {\bibfnamefont {X.-M.}\ \bibnamefont {Lu}},\ }in\ \href
  {\doibase 10.1117/12.2245733} {\emph {\bibinfo {booktitle} {Proc. SPIE}}},\
  \bibinfo {series} {Quantum and Nonlinear Optics IV}, Vol.\ \bibinfo {volume}
  {10029}\ (\bibinfo {year} {2016})\ pp.\ \bibinfo {pages}
  {1002903--1002903--7}\BibitemShut {NoStop}%
\bibitem [{\citenamefont {Nair}\ and\ \citenamefont
  {Tsang}(2016{\natexlab{a}})}]{nair16}%
  \BibitemOpen
  \bibfield  {author} {\bibinfo {author} {\bibfnamefont {R.}~\bibnamefont
  {Nair}}\ and\ \bibinfo {author} {\bibfnamefont {M.}~\bibnamefont {Tsang}},\
  }\href {\doibase 10.1364/OE.24.003684} {\bibfield  {journal} {\bibinfo
  {journal} {Opt. Express}\ }\textbf {\bibinfo {volume} {24}},\ \bibinfo
  {pages} {3684} (\bibinfo {year} {2016}{\natexlab{a}})}\BibitemShut {NoStop}%
\bibitem [{\citenamefont {Tang}\ \emph {et~al.}(2016)\citenamefont {Tang},
  \citenamefont {Durak},\ and\ \citenamefont {Ling}}]{SDL16}%
  \BibitemOpen
  \bibfield  {author} {\bibinfo {author} {\bibfnamefont {Z.~S.}\ \bibnamefont
  {Tang}}, \bibinfo {author} {\bibfnamefont {K.}~\bibnamefont {Durak}}, \ and\
  \bibinfo {author} {\bibfnamefont {A.}~\bibnamefont {Ling}},\ }\href {\doibase
  10.1364/OE.24.022004} {\bibfield  {journal} {\bibinfo  {journal} {Opt.
  Express}\ }\textbf {\bibinfo {volume} {24}},\ \bibinfo {pages} {22004}
  (\bibinfo {year} {2016})}\BibitemShut {NoStop}%
\bibitem [{\citenamefont {Yang}\ \emph {et~al.}(2016)\citenamefont {Yang},
  \citenamefont {Tashchilina}, \citenamefont {Moiseev}, \citenamefont {Simon},\
  and\ \citenamefont {Lvovsky}}]{YTM+16}%
  \BibitemOpen
  \bibfield  {author} {\bibinfo {author} {\bibfnamefont {F.}~\bibnamefont
  {Yang}}, \bibinfo {author} {\bibfnamefont {A.}~\bibnamefont {Tashchilina}},
  \bibinfo {author} {\bibfnamefont {E.~S.}\ \bibnamefont {Moiseev}}, \bibinfo
  {author} {\bibfnamefont {C.}~\bibnamefont {Simon}}, \ and\ \bibinfo {author}
  {\bibfnamefont {A.~I.}\ \bibnamefont {Lvovsky}},\ }\href {\doibase
  10.1364/OPTICA.3.001148} {\bibfield  {journal} {\bibinfo  {journal} {Optica}\
  }\textbf {\bibinfo {volume} {3}},\ \bibinfo {pages} {1148} (\bibinfo {year}
  {2016})}\BibitemShut {NoStop}%
\bibitem [{\citenamefont {Tham}\ \emph {et~al.}(2017)\citenamefont {Tham},
  \citenamefont {Ferretti},\ and\ \citenamefont {Steinberg}}]{TFS17}%
  \BibitemOpen
  \bibfield  {author} {\bibinfo {author} {\bibfnamefont {W.-K.}\ \bibnamefont
  {Tham}}, \bibinfo {author} {\bibfnamefont {H.}~\bibnamefont {Ferretti}}, \
  and\ \bibinfo {author} {\bibfnamefont {A.~M.}\ \bibnamefont {Steinberg}},\
  }\href {\doibase 10.1103/PhysRevLett.118.070801} {\bibfield  {journal}
  {\bibinfo  {journal} {Phys. Rev. Lett.}\ }\textbf {\bibinfo {volume} {118}},\
  \bibinfo {pages} {070801} (\bibinfo {year} {2017})}\BibitemShut {NoStop}%
\bibitem [{\citenamefont {Pa\'{u}r}\ \emph {et~al.}(2016)\citenamefont
  {Pa\'{u}r}, \citenamefont {Stoklasa}, \citenamefont {Hradil}, \citenamefont
  {S\'{a}nchez-Soto},\ and\ \citenamefont {Rehacek}}]{PSH+16}%
  \BibitemOpen
  \bibfield  {author} {\bibinfo {author} {\bibfnamefont {M.}~\bibnamefont
  {Pa\'{u}r}}, \bibinfo {author} {\bibfnamefont {B.}~\bibnamefont {Stoklasa}},
  \bibinfo {author} {\bibfnamefont {Z.}~\bibnamefont {Hradil}}, \bibinfo
  {author} {\bibfnamefont {L.~L.}\ \bibnamefont {S\'{a}nchez-Soto}}, \ and\
  \bibinfo {author} {\bibfnamefont {J.}~\bibnamefont {Rehacek}},\ }\href
  {\doibase 10.1364/OPTICA.3.001144} {\bibfield  {journal} {\bibinfo  {journal}
  {Optica}\ }\textbf {\bibinfo {volume} {3}},\ \bibinfo {pages} {1144}
  (\bibinfo {year} {2016})}\BibitemShut {NoStop}%
\bibitem [{\citenamefont {Nair}\ and\ \citenamefont
  {Tsang}(2016{\natexlab{b}})}]{NT16Gaussian}%
  \BibitemOpen
  \bibfield  {author} {\bibinfo {author} {\bibfnamefont {R.}~\bibnamefont
  {Nair}}\ and\ \bibinfo {author} {\bibfnamefont {M.}~\bibnamefont {Tsang}},\
  }\href {\doibase 10.1103/PhysRevLett.117.190801} {\bibfield  {journal}
  {\bibinfo  {journal} {Physical Review Letters}\ }\textbf {\bibinfo {volume}
  {117}},\ \bibinfo {pages} {190801} (\bibinfo {year}
  {2016}{\natexlab{b}})}\BibitemShut {NoStop}%
\bibitem [{\citenamefont {Lupo}\ and\ \citenamefont
  {Pirandola}(2016)}]{lupo16}%
  \BibitemOpen
  \bibfield  {author} {\bibinfo {author} {\bibfnamefont {C.}~\bibnamefont
  {Lupo}}\ and\ \bibinfo {author} {\bibfnamefont {S.}~\bibnamefont
  {Pirandola}},\ }\href {\doibase 10.1103/PhysRevLett.117.190802} {\bibfield
  {journal} {\bibinfo  {journal} {Phys. Rev. Lett.}\ }\textbf {\bibinfo
  {volume} {117}},\ \bibinfo {pages} {190802} (\bibinfo {year}
  {2016})}\BibitemShut {NoStop}%
\bibitem [{\citenamefont {{Rehacek}}\ \emph {et~al.}(2016)\citenamefont
  {{Rehacek}}, \citenamefont {{Paur}}, \citenamefont {{Stoklasa}},
  \citenamefont {{Motka}}, \citenamefont {{Hradil}},\ and\ \citenamefont
  {{Sanchez-Soto}}}]{RPS+16}%
  \BibitemOpen
  \bibfield  {author} {\bibinfo {author} {\bibfnamefont {J.}~\bibnamefont
  {{Rehacek}}}, \bibinfo {author} {\bibfnamefont {M.}~\bibnamefont {{Paur}}},
  \bibinfo {author} {\bibfnamefont {B.}~\bibnamefont {{Stoklasa}}}, \bibinfo
  {author} {\bibfnamefont {L.}~\bibnamefont {{Motka}}}, \bibinfo {author}
  {\bibfnamefont {Z.}~\bibnamefont {{Hradil}}}, \ and\ \bibinfo {author}
  {\bibfnamefont {L.~L.}\ \bibnamefont {{Sanchez-Soto}}},\ }\href@noop {}
  {\bibfield  {journal} {\bibinfo  {journal} {ArXiv e-prints}\ } (\bibinfo
  {year} {2016})},\ \Eprint {http://arxiv.org/abs/1607.05837} {arXiv:1607.05837
  [quant-ph]} \BibitemShut {NoStop}%
\bibitem [{\citenamefont {{Kerviche}}\ \emph {et~al.}(2017)\citenamefont
  {{Kerviche}}, \citenamefont {{Guha}},\ and\ \citenamefont {{Ashok}}}]{KGA17}%
  \BibitemOpen
  \bibfield  {author} {\bibinfo {author} {\bibfnamefont {R.}~\bibnamefont
  {{Kerviche}}}, \bibinfo {author} {\bibfnamefont {S.}~\bibnamefont {{Guha}}},
  \ and\ \bibinfo {author} {\bibfnamefont {A.}~\bibnamefont {{Ashok}}},\
  }\href@noop {} {\bibfield  {journal} {\bibinfo  {journal} {ArXiv e-prints}\ }
  (\bibinfo {year} {2017})},\ \Eprint {http://arxiv.org/abs/1701.04913}
  {arXiv:1701.04913 [physics.optics]} \BibitemShut {NoStop}%
\bibitem [{\citenamefont {Tsang}(2017)}]{Tsa17}%
  \BibitemOpen
  \bibfield  {author} {\bibinfo {author} {\bibfnamefont {M.}~\bibnamefont
  {Tsang}},\ }\href {http://stacks.iop.org/1367-2630/19/i=2/a=023054}
  {\bibfield  {journal} {\bibinfo  {journal} {New Journal of Physics}\ }\textbf
  {\bibinfo {volume} {19}},\ \bibinfo {pages} {023054} (\bibinfo {year}
  {2017})}\BibitemShut {NoStop}%
\bibitem [{\citenamefont {Lu}\ \emph {et~al.}(2016)\citenamefont {Lu},
  \citenamefont {Nair},\ and\ \citenamefont {Tsang}}]{lu16}%
  \BibitemOpen
  \bibfield  {author} {\bibinfo {author} {\bibfnamefont {X.-M.}\ \bibnamefont
  {Lu}}, \bibinfo {author} {\bibfnamefont {R.}~\bibnamefont {Nair}}, \ and\
  \bibinfo {author} {\bibfnamefont {M.}~\bibnamefont {Tsang}},\ }\href
  {http://arxiv.org/abs/1609.03025} {\bibfield  {journal} {\bibinfo  {journal}
  {arXiv:1609.03025 [quant-ph]}\ } (\bibinfo {year} {2016})},\ \bibinfo {note}
  {arXiv: 1609.03025}\BibitemShut {NoStop}%
\bibitem [{\citenamefont {{Krovi}}\ \emph {et~al.}(2016)\citenamefont
  {{Krovi}}, \citenamefont {{Guha}},\ and\ \citenamefont {{Shapiro}}}]{KGS16}%
  \BibitemOpen
  \bibfield  {author} {\bibinfo {author} {\bibfnamefont {H.}~\bibnamefont
  {{Krovi}}}, \bibinfo {author} {\bibfnamefont {S.}~\bibnamefont {{Guha}}}, \
  and\ \bibinfo {author} {\bibfnamefont {J.~H.}\ \bibnamefont {{Shapiro}}},\
  }\href@noop {} {\bibfield  {journal} {\bibinfo  {journal} {ArXiv e-prints}\ }
  (\bibinfo {year} {2016})},\ \Eprint {http://arxiv.org/abs/1609.00684}
  {arXiv:1609.00684 [quant-ph]} \BibitemShut {NoStop}%
\bibitem [{\citenamefont {Hayashi}(2005)}]{hayashi05}%
  \BibitemOpen
  \bibfield  {author} {\bibinfo {author} {\bibfnamefont {M.}~\bibnamefont
  {Hayashi}},\ }\href {https://books.google.com.sg/books?id=jYwrP1mQR\_gC}
  {\emph {\bibinfo {title} {Asymptotic Theory of Quantum Statistical Inference:
  Selected Papers}}}\ (\bibinfo  {publisher} {World Scientific},\ \bibinfo
  {year} {2005})\BibitemShut {NoStop}%
\bibitem [{\citenamefont {Fujiwara}(2006)}]{fujiwara06}%
  \BibitemOpen
  \bibfield  {author} {\bibinfo {author} {\bibfnamefont {A.}~\bibnamefont
  {Fujiwara}},\ }\href {http://stacks.iop.org/0305-4470/39/i=40/a=014}
  {\bibfield  {journal} {\bibinfo  {journal} {Journal of Physics A:
  Mathematical and General}\ }\textbf {\bibinfo {volume} {39}},\ \bibinfo
  {pages} {12489} (\bibinfo {year} {2006})}\BibitemShut {NoStop}%
\bibitem [{\citenamefont {Goodman}(2005)}]{goodman05}%
  \BibitemOpen
  \bibfield  {author} {\bibinfo {author} {\bibfnamefont {J.}~\bibnamefont
  {Goodman}},\ }\href@noop {} {\emph {\bibinfo {title} {Introduction to Fourier
  Optics}}},\ \bibinfo {edition} {3rd}\ ed.\ (\bibinfo  {publisher} {Roberts \&
  Company Publishers},\ \bibinfo {year} {2005})\BibitemShut {NoStop}%
\bibitem [{\citenamefont {Wiseman}\ and\ \citenamefont
  {Milburn}(2010)}]{wiseman10}%
  \BibitemOpen
  \bibfield  {author} {\bibinfo {author} {\bibfnamefont {H.}~\bibnamefont
  {Wiseman}}\ and\ \bibinfo {author} {\bibfnamefont {G.}~\bibnamefont
  {Milburn}},\ }\href@noop {} {\emph {\bibinfo {title} {Quantum Measurement and
  Control}}}\ (\bibinfo  {publisher} {Cambridge University Press},\ \bibinfo
  {year} {2010})\BibitemShut {NoStop}%
\bibitem [{\citenamefont {Braunstein}\ and\ \citenamefont
  {Caves}(1994)}]{BC94}%
  \BibitemOpen
  \bibfield  {author} {\bibinfo {author} {\bibfnamefont {S.~L.}\ \bibnamefont
  {Braunstein}}\ and\ \bibinfo {author} {\bibfnamefont {C.~M.}\ \bibnamefont
  {Caves}},\ }\href {\doibase 10.1103/PhysRevLett.72.3439} {\bibfield
  {journal} {\bibinfo  {journal} {Physical Review Letters}\ }\textbf {\bibinfo
  {volume} {72}},\ \bibinfo {pages} {3439} (\bibinfo {year}
  {1994})}\BibitemShut {NoStop}%
\bibitem [{\citenamefont {Mandel}\ and\ \citenamefont {Wolf}(1995)}]{mandel95}%
  \BibitemOpen
  \bibfield  {author} {\bibinfo {author} {\bibfnamefont {L.}~\bibnamefont
  {Mandel}}\ and\ \bibinfo {author} {\bibfnamefont {E.}~\bibnamefont {Wolf}},\
  }\href {https://books.google.com.sg/books?id=FeBix14iM70C} {\emph {\bibinfo
  {title} {Optical Coherence and Quantum Optics}}}\ (\bibinfo  {publisher}
  {Cambridge University Press},\ \bibinfo {year} {1995})\BibitemShut {NoStop}%
\bibitem [{\citenamefont {Shapiro}(2009)}]{Sha09}%
  \BibitemOpen
  \bibfield  {author} {\bibinfo {author} {\bibfnamefont {J.~H.}\ \bibnamefont
  {Shapiro}},\ }\href {\doibase 10.1109/JSTQE.2009.2024959} {\bibfield
  {journal} {\bibinfo  {journal} {IEEE Journal of Selected Topics in Quantum
  Electronics}\ }\textbf {\bibinfo {volume} {15}},\ \bibinfo {pages} {1547 }
  (\bibinfo {year} {2009})}\BibitemShut {NoStop}%
\bibitem [{\citenamefont {Gerry}\ and\ \citenamefont
  {Knight}(2005)}]{GK05quantum}%
  \BibitemOpen
  \bibfield  {author} {\bibinfo {author} {\bibfnamefont {C.}~\bibnamefont
  {Gerry}}\ and\ \bibinfo {author} {\bibfnamefont {P.}~\bibnamefont {Knight}},\
  }\href@noop {} {\emph {\bibinfo {title} {Introductory Quantum Optics}}}\
  (\bibinfo  {publisher} {Cambridge University Press},\ \bibinfo {year}
  {2005})\BibitemShut {NoStop}%
\bibitem [{\citenamefont {Yariv}(1989)}]{yariv89}%
  \BibitemOpen
  \bibfield  {author} {\bibinfo {author} {\bibfnamefont {A.}~\bibnamefont
  {Yariv}},\ }\href@noop {} {\emph {\bibinfo {title} {Quantum Electronics}}}\
  (\bibinfo  {publisher} {Wiley, New York},\ \bibinfo {year}
  {1989})\BibitemShut {NoStop}%
\bibitem [{\citenamefont {Zhang}\ and\ \citenamefont {Li}(2013)}]{zhang13}%
  \BibitemOpen
  \bibfield  {author} {\bibinfo {author} {\bibfnamefont {K.}~\bibnamefont
  {Zhang}}\ and\ \bibinfo {author} {\bibfnamefont {D.}~\bibnamefont {Li}},\
  }\href {https://books.google.com.sg/books?id=7DnvCAAAQBAJ} {\emph {\bibinfo
  {title} {Electromagnetic Theory for Microwaves and Optoelectronics}}}\
  (\bibinfo  {publisher} {Springer Berlin Heidelberg},\ \bibinfo {year}
  {2013})\BibitemShut {NoStop}%
\bibitem [{\citenamefont {{Tsang}}(2016)}]{Tsa16}%
  \BibitemOpen
  \bibfield  {author} {\bibinfo {author} {\bibfnamefont {M.}~\bibnamefont
  {{Tsang}}},\ }\href@noop {} {\  (\bibinfo {year} {2016})},\ \Eprint
  {http://arxiv.org/abs/1605.03799} {arXiv:1605.03799 [quant-ph]} \BibitemShut
  {NoStop}%
\bibitem [{\citenamefont {Wicker}\ \emph {et~al.}(2009)\citenamefont {Wicker},
  \citenamefont {Sindbert},\ and\ \citenamefont {Heintzmann}}]{WSH09}%
  \BibitemOpen
  \bibfield  {author} {\bibinfo {author} {\bibfnamefont {K.}~\bibnamefont
  {Wicker}}, \bibinfo {author} {\bibfnamefont {S.}~\bibnamefont {Sindbert}}, \
  and\ \bibinfo {author} {\bibfnamefont {R.}~\bibnamefont {Heintzmann}},\
  }\href {\doibase 10.1364/OE.17.015491} {\bibfield  {journal} {\bibinfo
  {journal} {Optics Express}\ }\textbf {\bibinfo {volume} {17}},\ \bibinfo
  {pages} {15491} (\bibinfo {year} {2009})}\BibitemShut {NoStop}%
\bibitem [{\citenamefont {Weigel}\ \emph {et~al.}(2011)\citenamefont {Weigel},
  \citenamefont {Babovsky}, \citenamefont {Kiessling},\ and\ \citenamefont
  {Kowarschik}}]{WBK+11}%
  \BibitemOpen
  \bibfield  {author} {\bibinfo {author} {\bibfnamefont {D.}~\bibnamefont
  {Weigel}}, \bibinfo {author} {\bibfnamefont {H.}~\bibnamefont {Babovsky}},
  \bibinfo {author} {\bibfnamefont {A.}~\bibnamefont {Kiessling}}, \ and\
  \bibinfo {author} {\bibfnamefont {R.}~\bibnamefont {Kowarschik}},\ }\href
  {\doibase 10.1016/j.optcom.2010.12.068} {\bibfield  {journal} {\bibinfo
  {journal} {Optics Communications}\ }\textbf {\bibinfo {volume} {284}},\
  \bibinfo {pages} {2273} (\bibinfo {year} {2011})}\BibitemShut {NoStop}%
\end{thebibliography}
\end{document}